\documentclass[useAMS]{mn2e}
\usepackage{graphicx}
\usepackage{psfig}
\def\lsim{\,\lower2truept\hbox{${<\atop\hbox{\raise4truept\hbox{$\sim$}}}$}\,}
\def\gsim{\,\lower2truept\hbox{${> \atop\hbox{\raise4truept\hbox{$\sim$}}}$}\,}

\title[Large-scale sub-mm surveys]{Astrophysical and Cosmological
Information from Large-scale sub-mm Surveys of Extragalactic
Sources}
\author[M. Negrello et al. ]{M. Negrello$^{1,2}$\thanks{E-mail:
M.Negrello@open.ac.uk}, F. Perrotta$^{1,3}$,  J. Gonz\'alez-Nuevo
Gonz\'alez$^{1}$, L. Silva$^{3}$, \newauthor 
              G. De Zotti$^{4,1}$, G.L.
Granato$^{4,1}$, C. Baccigalupi$^{1}$, and L. Danese$^{1}$ \\
\ \\
$^{1}$SISSA, Via Beirut 4, I-34014, Trieste, Italy \\
$^{2}$Department of Physics and Astronomy, Open University, Walton
Hall, Milton Keynes MK7 6AA, UK \\
$^{3}${INAF - Osservatorio Astronomico di Trieste, Via G.B.
Tiepolo 11, I-34131 Trieste, Italy}\\
$^{4}$INAF -- Osservatorio Astronomico di Padova, Vicolo
dell'Osservatorio 5, I-35122 Padova, Italy }
\begin{document}

\date{....}

\pagerange{\pageref{firstpage}--\pageref{lastpage}} \pubyear{2006}

\maketitle

\label{firstpage}

\begin{abstract}
We present a quantitative analysis of the astrophysical and
cosmological information that can be extracted from the many
important wide-area, shallow surveys that will be carried out in
the next few years. Our calculations combine the predictions of
the physical model by Granato et al. (2004) for the formation and
evolution of spheroidal galaxies with up-to-date phenomenological
models for the evolution of starburst and normal late-type
galaxies and of radio sources. We compute the expected number
counts and the redshift distributions of these source populations
separately and then focus on proto-spheroidal galaxies. For the
latter objects we predict the counts and redshift distributions of
strongly lensed sources at 250, 350, 500, and $850\,\mu$m, the
angular correlation function of sources detected in the surveys
considered, the angular power spectra due to clustering of sources
below the detection limit in {\sc Herschel} and {\sc Planck}
surveys. An optimal survey for selecting strongly lensed
proto-spheroidal galaxies is described, and it is shown how they
can be easily distinguished from the other source populations. We
also discuss the detectability of the imprints of the 1-halo and
2-halo regimes on angular correlation functions and clustering
power spectra, as well as the constraints on cosmological
parameters that can be obtained from the determinations of these
quantities. The novel data relevant to derive the first
sub-millimeter estimates of the local luminosity functions of
starburst and late-type galaxies, and the constraints on the
properties of rare source populations, such as blazars, are also
briefly described.

\end{abstract}


\section{Introduction}

So far, surveys at sub-millimeter (sub-mm) wavelengths have
covered only tiny fractions of the sky (less than
$1\,\hbox{deg}^2$) down to mJy levels (Mortier et al. 2005, and
references therein). Forthcoming experiments, however, will
enormously extend the sky coverage. The Legacy Surveys with
SCUBA-2\footnote{ www.roe.ac.uk/ukatc/projects/scubatwo/} include
a Shallow Survey (SASSy) of $\sim 20,000\,\hbox{deg}^2$ to a depth
of 150 mJy ($5\sigma$) and a Wide-Area Extragalactic
Survey\footnote{After this paper was submitted, we learned that
the Wide-Area Extragalactic Survey at $850\mu$m has been combined
with the Deep $450\mu$m survey. The joint programme, known as the
SCUBA-2 Cosmology Legacy Survey
(www.strw.leidenuniv.nl/$\sim$pvdwerf/pdf/SCUBA2-CLS.pdf), has two
elements: an $850\mu$m survey of $\sim 35\,\hbox{deg}^2$ to a
$5\sigma$ limit of 3.5 mJy and a deeper $450\mu$m survey of $\sim
1.5\,\hbox{deg}^2$ to a $5\sigma$ limit of 2.5 mJy.} of
$5\,\hbox{deg}^2$ to a $5\sigma$ flux limit of 2 mJy, both at
$850\mu$m. The SPIRE instrument on ESA's {\sc Herschel} satellite
may be able to carry out a shallow survey of $\sim
400\,\hbox{deg}^2$ to a 100 mJy level ($5\sigma$) in all three
bands (250, 350, and $500\mu$m; Lagache et al. 2003; Harwit 2004).
The HFI instrument (Lamarre et al. 2003) on the {\sc Planck}
satellite will carry out all sky surveys in 6 bands, centered at
3000, 2100, 1380, 850, 550, and 350 $\mu$m; these surveys will be
confusion limited to flux levels ranging from a few hundred mJy to
$\simeq 1.5\,$Jy (Vielva et al. 2003; Negrello et al. 2004;
L{\'o}pez-Caniego et al. 2006).

As first pointed out by Blain (1996), the sub-mm region is
exceptionally well suited to exploit strong gravitational lensing
to detect large complete samples of high-redshift dusty galaxies.
The reason resides in the combined effect of the large and
negative K-correction, that makes the observed flux density of
sources of given luminosity only weakly dependent on distance over
a broad redshift range above $z\gsim 1$ (Blain \& Longair 1993),
and of very steep evolution. These two ingredients cooperate in
yielding large lensing optical depths and extremely steep counts,
resulting in a strong magnification bias.

Strong lensing is a powerful tool for many interesting
applications (see Kochanek 2004 for a comprehensive review). The
statistics of lenses is a test of the cosmological model because
the optical depth is roughly proportional to the volume to the
source. If the redshifts of both the source and the lens are
measured, the ratio between the lens-source and observer-source
angular diameter distances, $d_{ls}/d_{os}$, constitutes a
sensitive probe of the geometry of the Universe (and in particular
of $\Omega_\Lambda$, with weak dependence on $\Omega_m$) through
the scaling of light-ray deflections with the source redshift. To
date, in the best studied cases, the lens is a low redshift
cluster such as A1689 or A2218, and this test is difficult to
apply because, for high redshift sources, $d_{ls}/d_{os}$ is
always $\simeq 1$ and the effect of different cosmologies is very
small. On the contrary, in the case of the sub-mm selected
sources, the lenses are expected to be generally at substantial
redshifts. The distribution of lens redshifts provides a further,
potentially powerful, cosmological test since, in general,
cosmologies with a large cosmological constant predict
significantly higher lens redshifts than those without. Recent
analyses of lens statistics have focussed on the CLASS survey (see
e.g., Mitchell et al. 2005) to derive constraints on cosmological
parameters. Attempts to use the lens statistics to constrain dark
energy have been worked out by Chae et al. (2004) and Kuhlen et
al. (2004).

On the other hand, the probability that a source has an
intervening lens also depends on the distribution of the lens
galaxies, and the cross section of the lens for producing a given
effect depends on its velocity dispersion. Thus, if a cosmological
model is adopted, the lens statistics is informative on the
evolution of galaxy properties, and in particular of dark matter
structures. Note that gravitational lenses are unique in providing
a selection based on mass, rather than luminosity, so that the
separation distribution of lenses is a direct mapping of the mass
function of all halos, luminous or dark.

Alexander et al. (2003, 2005) have found that a substantial
fraction of sub-mm bright galaxies harbor an active galactic
nucleus (AGN). The magnification produced by gravitational lensing
allows us to study fainter AGN host galaxies than otherwise
possible. This will permit a better understanding of the growth of
super-massive black holes and their relationships with their host
galaxies.

Another important driver of large-area sub-mm surveys is the
investigation of the clustering properties of high-$z$ galaxies,
both in the 1-halo and in the 2-halo regimes, that can be
performed more effectively than in other wavebands, and without
being affected by obscuration effects. The large scale angular
correlation function of detected sources and the autocorrelation
function of intensity fluctuations provide information on the
masses of dark matter halos, on the duration of the star-formation
process and on the evolution of the bias factor. They are also
sensitive to, and will provide constraints on cosmological
parameters.

Furthermore, these large area surveys will yield complete samples
of sub-mm selected low-$z$ dusty galaxies, allowing, for the first
time, direct estimates of the local luminosity functions.  In
addition, they may contain a number of extreme luminosity distant
galaxies.

Another interesting, although only seldom mentioned, product of
large-area sub-mm surveys are complete samples of radio sources,
selected in a spectral region almost unexplored so far, but where
key information on their physics is expected to show up. For
example, observations at these wavelengths may reveal the
transition from optically thick to optically thin synchrotron
emission in the most compact regions, allowing the determination
of the self-absorption (turnover) frequency which carries
information on physical parameters.  Also, the slope of the
optically thin synchrotron emission steepens at high frequencies
due to electron energy losses; the spectral break frequency
$\nu_b$ is related to the strength of the magnetic field and to
the `synchrotron age' $t_s$ by $\nu_b \simeq
320(20\mu\hbox{G}/\hbox{B})^3 (t_s/\hbox{Myr})^{-2}\,$GHz. Excess
far-IR/sub-mm emission possibly due to dust, is often observed
from radio galaxies (Knapp \& Patten 1991). These surveys will
allow us to assess whether this is a general property of these
sources, with interesting implications for the presence of dust in
the host galaxies, generally identified with giant ellipticals,
usually devoid of interstellar matter.

The dominant radio source population at the relevant frequencies
is made of flat-spectrum radio quasars and of BL Lac objects,
collectively dubbed `blazars'. Multifrequency surveys in the
far-IR/sub-mm region will allow us to check, e.g., whether there
are systematic differences in the turnover frequencies between BL
Lacs and flat-spectrum quasars as expected if the BL Lac emission
is angled closer to the line of sight, so that the turnovers are
Doppler boosted to higher frequencies.

Correlations between turnover frequency and luminosity, which is
also boosted by relativistic beaming effects, may be expected. On
the other hand, Fossati et al. (1998) and Ghisellini et al. (1998)
found an anti-correlation between the frequency of the synchrotron
peak of blazars and their radio luminosity, referred to as `the
blazar sequence'. The spectral energy distributions of the most
luminous blazars appear to peak at (rest-frame) mm wavelengths,
while the peak moves towards optical wavelengths as the luminosity
decreases. This sequence, if confirmed, is an important
breakthrough in our understanding of the physical processes
governing the blazar emission. On the other hand, Ant\'on \&
Browne (2005) find that most of their low-luminosity flat-spectrum
radio sources peak at much lower frequencies than expected from
the Fossati et al. (1998) relationship. Also, Nieppola et al.
(2006) did not find any correlation between luminosity and
frequency of the synchrotron peak for their sample of 300 blazars.
However, only a limited region of the luminosity-peak frequency
plane has been explored to date, mostly due to the dearth of data
in the far-IR to mm region, and the samples have been collected
from different surveys and are not complete.

In this paper we address all these important aspects. After an
overview of the expected counts of extragalactic sources at sub-mm
wavelengths (\S\,2), we present a comparative study of the
potential of these surveys in relation to: the selection of
strongly lensed, dusty primordial galaxies (\S\,3), the clustering
properties of sub-mm sources at high redshifts (\S\,4), and the
selection of the first complete sub-mm samples of local
star-forming galaxies and of radio sources, mainly blazars
(\S\,5). In \S\,6 we summarize and discuss our main conclusions.
This investigation may help in guiding us on what to expect and in
optimizing the survey programmes.

Throughout this paper we will adopt the ``concordance model'',
i.e. a flat $\Lambda$CDM cosmology with $\Omega_m=0.27$,
$\Omega_b=0.045$, and $\Omega_\Lambda=0.73$, $H_0=72{\,\rm
km~s^{-1}~Mpc^{-1}}$, $\sigma_8=0.8$, and an index $n=1.0$ for the
power spectrum of primordial density fluctuations.

\begin{figure}
\includegraphics[width=0.9\columnwidth]{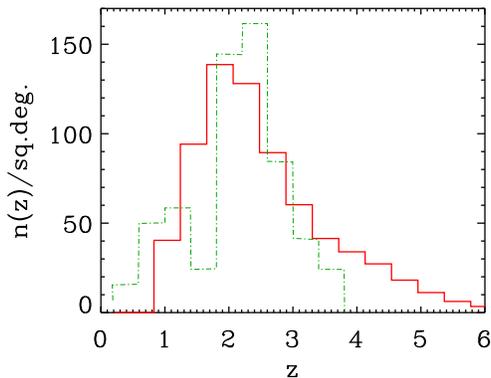}
\caption{Comparison of the redshift distribution of sources with
$S_{850\mu{\rm m}} > 5\,$mJy yielded by the Granato et al. (2004)
model (solid histogram) with the observational estimate by Chapman
et al. (2005; dot-dashed). The numbers of sources per bin of
Chapman et al. (2005) have been rescaled to a surface density
$N(S_{850\mu{\rm m}}>5\,\hbox{mJy})= 620\,\hbox{deg}^{-2}$ (Scott
et al. 2002). As pointed out by Chapman et al. (2005), the dip at
$z\simeq 1.5$ and the high-$z$ cut-off in the data are likely due
to the radio and optical selections. } \label{850nz}
\end{figure}

\begin{figure}
\centering
\includegraphics[width=1.0\columnwidth]{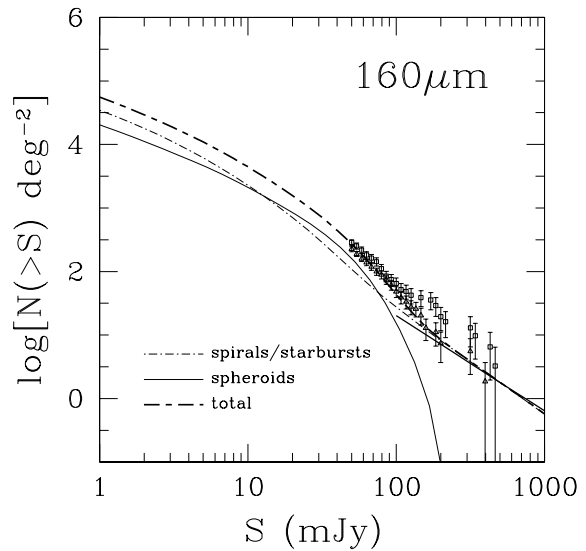}
\caption{Model $160\,\mu$m integral counts compared with the
Spitzer counts by Dole et al. (2004; squares) and Frayer et al.
(2006a; triangles). The heavy solid line on the lower left-hand
corner shows the count estimate by Serjeant \& Harrison (2005). }
\label{c160}
\end{figure}

\begin{figure*}
\includegraphics[width=0.85\textwidth]{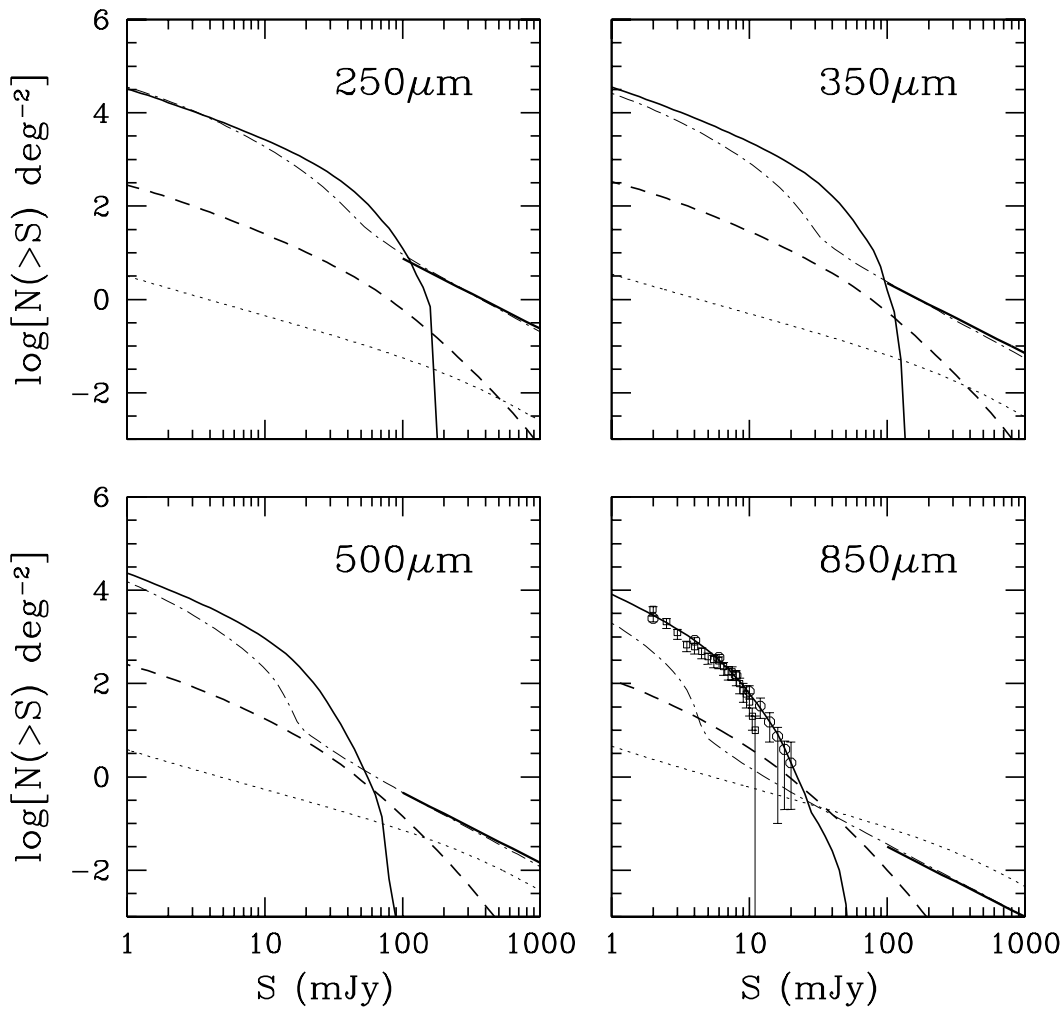}
\label{counts} \caption{Predicted counts of high-$z$
proto-spheroidal galaxies (solid line), late-type (starburst plus
normal spiral) galaxies (dot-dashed line), and radio sources
(dotted line). The dashed line shows the predicted counts of
strongly lensed proto-spheroidal galaxies. The $850\,\mu$m data
are from Coppin et al. (2006; circles) and Scott et al. (2006;
squares). The heavy solid line on the lower right-hand corner of
each panel shows the count estimates by Serjeant \& Harrison
(2005).} \label{fig:counts}
\end{figure*}

\section{Overview of the models and sub-mm counts}

Sub-millimeter extragalactic sources are a mixed bag of various
populations of dusty galaxies and of flat-spectrum radio sources.
A summary of the models adopted for each population follows.

\subsection{Proto-spheroidal galaxies}

Dusty galaxies detected at mJy levels by SCUBA and MAMBO surveys
are modelled following Granato et al. (2004), who interpret them
as proto-spheroidal galaxies in the process of forming most of
their stars in a gigantic starburst lasting 0.3 -- 1 Gyr
(depending on redshift and on the halo mass). This physical model
predicts the birth rate of galaxies as a function of cosmic time
and the evolution with galactic age of their full spectral energy
distribution (SED), from radio to X-ray wavelengths, including the
effect of the active nucleus, whose growth is closely tied to the
evolution of the host galaxy. It is thus strongly constrained by,
and indeed accounts for huge amounts of multifrequency data at
different redshifts (Silva et al. 2005; Cirasuolo et al. 2005;
Shankar et al. 2006; Granato et al. 2006; Lapi et al. 2006). In
particular, it fits the $850\,\mu$m counts available at the time
(Granato et al. 2004) and the data on the redshift distribution
(Fig.~\ref{850nz}).

However, the re-analysis of existing blank-field SCUBA surveys by
Scott et al. (2006) and the results of the SCUBA Half-Degree
Extragalactic Survey (Coppin et al. 2006) converge in yielding
lower counts than previous studies. To match these more recent
results, we have scaled the $850\,\mu$m fluxes of Granato et al.
(2004) by a factor of 0.68, thus preserving the agreement with the
shape of the redshift distribution.

Most recently, Aretxaga et al. (2007) compared the redshift
distribution of the SCUBA Half Degree Extragalactic Survey
(SHADES), derived from rest-frame radio-mm-far IR colours, with
those yielded by the four different galaxy formation models
studied by van Kampen et al. (2005) and with the semi-analytic
models by Baugh et al. (2005) and by Granato et al. (2004). Only
the latter model turned out to be compatible, within the
uncertainties, with the data, although if sources detected only at
$850\,\mu$m are introduced in the redshift probability
distribution with other priors, two of the models considered by
van Kampen et al. (2005) can come closer to acceptable.

\subsection{Starburst and normal disk galaxies}

The Granato et al. (2004) model is meant to take into account the
star formation occurring within galactic dark-matter halos
virialized at $z_{\rm vir} \gsim 1.5$ and bigger than $M_{\rm vir}
\simeq 10^{11.6} M_\odot$, which are, crudely, associated to
spheroidal galaxies. We envisage disks (and irregulars) as
associated primarily to halos virializing at $z_{\rm vir} \lsim
1.5$, which have incorporated, through merging processes, a large
fraction of halos less massive than $4 \times 10^{11}\,M_\odot$
virializing at earlier times, which may become the bulges of late
type galaxies.

The contributions of the latter galaxies to the number counts are
estimated following, as usual, a phenomenological approach, which
consists in simple analytic recipes to evolve their local
luminosity functions (LFs), and appropriate templates for their
SEDs to compute K-corrections and to extrapolate the models to
different wavelengths where direct determinations of the local
luminosity functions are not available. We have adopted the
$60\,\mu$m local LFs by Saunders et al. (1990), that have been
derived for ``warm'' and ``cold'' galaxies (based on IRAS colors)
that we associate, respectively, to starburst and spiral galaxies.
Transformations to longer wavelengths were done using the mean
flux ratios determined by Serjeant \& Harrison (2005) who
extrapolated the IRAS detections using the sub-mm colour
temperature relations from the SCUBA Local Universe Survey (Dunne
et al. 2000; Dunne \& Eales 2001).

Rather detailed information on dusty galaxies at redshifts $\le
1.5$ came from the ISOCAM cosmological surveys performed with the
LW3 filter, centered at $15\,\mu$m (see Lagache et al. 2005 for a
review). Various phenomenological models proposed to account for
these data have been compared and discussed by Gruppioni et al.
(2002), who showed that the data require a combination of
luminosity and density evolution for starburst galaxies, stopping
at some redshift $z_{\rm break}\sim 1$. Gruppioni et al. (2002)
also allowed for non-evolving normal spirals and for type 1 active
galactic nuclei evolving in luminosity. The latter population,
however, yields a very small contribution to the $15\,\mu$m counts
(as well as to the sub-mm counts) and can be neglected here. On
the other hand, chemo/spectrophotometric evolution models for
normal disc galaxies (Mazzei et al. 1992; Matteucci 2006) indicate
a mild increase of the star formation rate, hence of IR
luminosity, with look-back time. In view of that, Silva et al.
(2004) updated the Gruppioni et al. (2002) model by allowing for a
moderate luminosity evolution of normal spiral galaxies. They
obtained good fits to the observed $15\,\mu$m counts and redshift
distributions with the following recipe: i) the LF, $\Phi_{\rm
sb}$, of the starburst population evolves both in density and in
luminosity as $\Phi_{\rm sb}[L(z),z] = \Phi_{\rm
sb}[L(z)/(1+z)^{2.5}, z=0]\times (1+z)^{3.5}$, with a
high-luminosity cutoff of the local LF at $L_{60\mu{\rm m}} =
2\times 10^{32}$ erg/s/Hz; ii) spirals evolve in luminosity as
$L(z)=L(0)\,(1+z)^{1.5}$; iii) the evolution of both starburst and
spiral galaxies stops at $z_{\rm break}=1$ and the LFs keep
constant afterwards, up to a redshift cut-off $z_{\rm cutoff}
=1.5$, above which essentially all the massive halos are
associated, according to Granato et al. (2004), with spheroidal
galaxies. We have adopted this model.

As shown by Silva et al. (2004, 2005), the model accounts for a
broad variety of data, including optical, radio, near-, mid-, and
far-IR counts, and redshift distributions. Moreover the model
predicted, at the flux limit ($F_{24\rm \mu m}\ge 83\,\mu$Jy) of
the Chandra Deep Field South MIPS survey (Papovich et al. 2004), a
surface density of sources at $z \ge 1.5$ of $\simeq
1\,\hbox{arcmin}^2$, i.e. amounting to $\simeq 22\%$ of the
observed total surface density. This prediction was at variance
with those of major phenomenological models which, for that flux
limit, yielded (see Fig.~2 of P\'erez-Gonz\'alez et al. 2005 and
Fig.~6 of Caputi et al. 2006) either very few (Chary et al. 2004)
or almost 50\% (Lagache et al. 2004) sources at $z\simeq 1.5$. The
redshift distribution observationally determined by
P\'erez-Gonz\'alez et al. (2005) and Caputi et al. (2006) for
$F_{24\mu\rm m} \ge 83\,\mu$Jy has 24--28\% of sources at $z\ge
1.5$, in nice agreement with the Silva et al. (2004) model. We add
here (Fig.~\ref{c160}) a comparison with {\it Spitzer} counts at
$160\,\mu$m (Dole et al. 2004; Frayer et al. 2006a), the
wavelength closest to those relevant here, confirming once more
the consistency of the model with the data.

\subsection{Extragalactic radio sources}

For radio sources we have adopted the model by De Zotti et al.
(2005). In the wavelength range and at the relatively bright flux
densities  considered in this paper, the dominant population are
blazars (flat-spectrum radio quasars and BL Lac objects). The
usual power-law approximation of the spectra of these objects may
break down at sub-mm wavelengths, where the synchrotron emission
of even the most compact regions turns from the optically thick to
the optically thin regime and is further steepened by electron
ageing effects. To allow for the corresponding steepening of radio
spectra, we have adopted the spectral shape given by Donato et al.
(2001) and Costamante \& Ghisellini (2002), including their
(controversial, see Ant{\' o}n \& Browne 2005 and Nieppola et al.
2006) anti-correlation between break frequency and radio
luminosity (see \S$\,$\ref{blazar}).

\subsection{Sub-mm counts}

Figure~\ref{fig:counts} summarizes the expected contributions of
proto-spheroids, of late-type (starburst plus spiral) galaxies,
and of radio sources to the counts at 850, 550, 350, and 250
$\mu$m, yielded by the models described above. The $850\,\mu$m
panel shows that the SCUBA counts are almost entirely accounted
for by proto-spheroidal galaxies, with a minor contribution from
starburst galaxies, which, however, become increasingly important
at low flux densities. At bright fluxes, the counts of dusty
galaxies match the extrapolation of IRAS data by Serjeant \&
Harrison (2005). At the brightest levels, however, the counts may
be dominated by flat-spectrum radio sources.

\begin{figure}
\includegraphics[height=8cm, width=8cm]{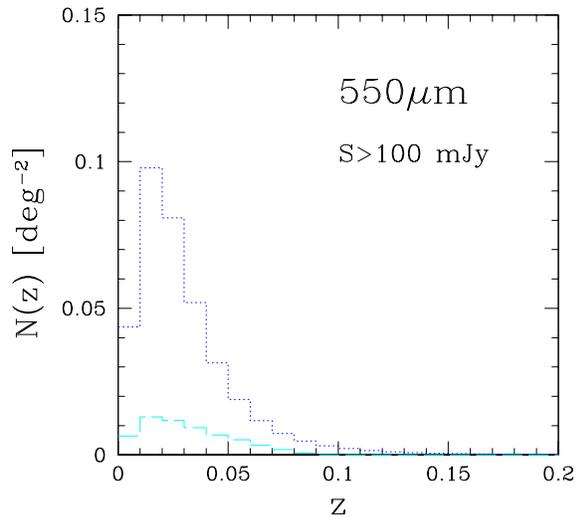} 
\caption{Redshift distribution of spiral (dotted) and starburst
(dashed) galaxies brighter than 100 mJy at $550\,\mu$m, for
$\Delta z=0.01$, according to the model described in the text.
}\label{sbspzdistr}
\end{figure}

\begin{figure*}
\includegraphics[width=0.90\textwidth]{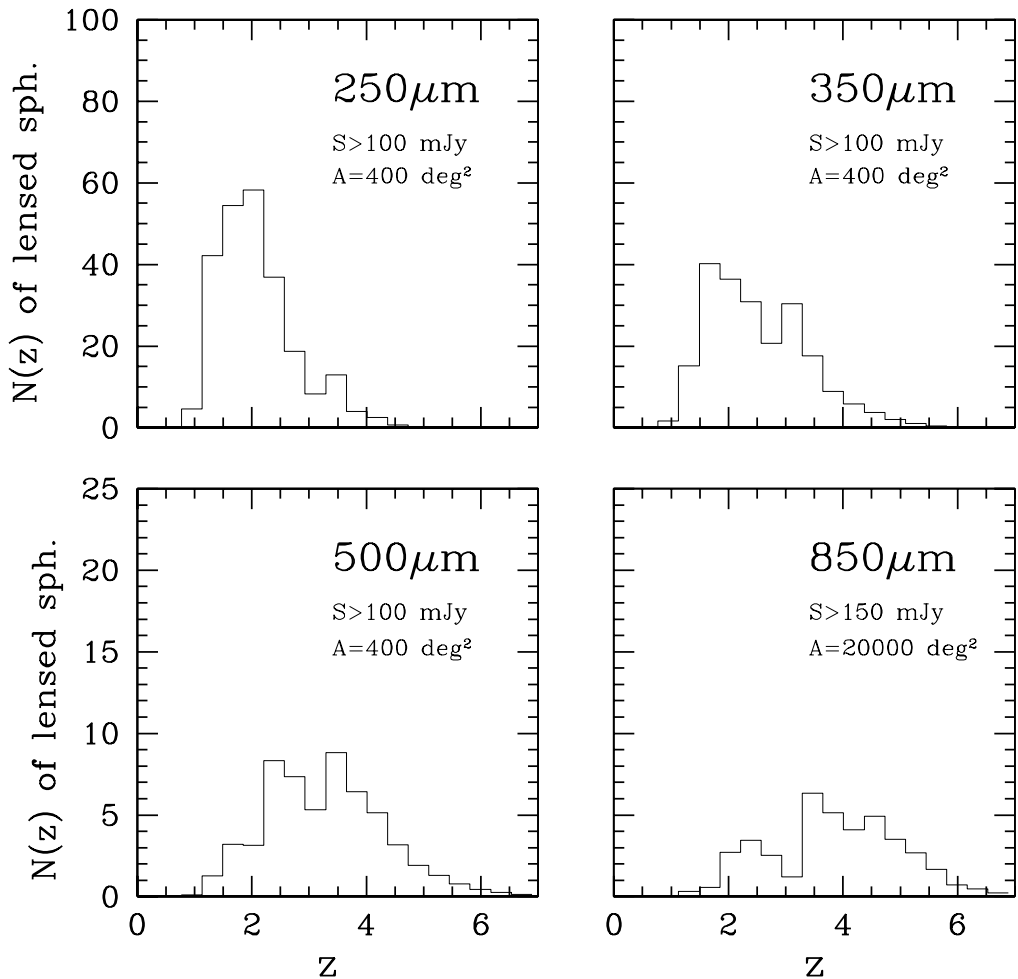}
\vskip-1.5truecm \caption{Predicted redshift distributions in bins
of $\Delta z=0.36$ of strongly lensed galaxies brighter than 100
mJy in the 3 {\sc Herschel}/SPIRE channels (250, 350, and
$550\,\mu$m) over an area of $400\,\hbox{deg}^2$, and brighter
that 150 mJy (the flux limit of the SCUBA-2 SASSy survey) at
$850\,\mu$m over an area of $20,000\,\hbox{deg}^2$, based on the
Granato et al. (2004) model, updated as described in the text. }
\label{zdistr}
\end{figure*}

\section{Strongly lensed proto-spheroidal
galaxies}\label{sec:lensing}

As already mentioned, the combination of the very steep counts
(Fig.~\ref{fig:counts}) and high redshifts (Fig.~\ref{850nz}),
hence large lensing optical depths, of sources detected by SCUBA
surveys maximizes the fraction of strongly lensed sources.

Calculations have been carried out as in Perrotta et al. (2002,
2003), but with the evolutionary model for proto-spheroidal
galaxies by Granato et al. (2004), updated as described above. The
predicted counts of strongly lensed sources (defined as those with
a gravitational amplification by a factor $\ge 2$) are displayed
in Fig.~\ref{fig:counts}. We find that the fraction of strongly
lensed proto-spheroids can indeed be very high at relatively
bright sub-mm fluxes, where the surface density of unlensed
proto-spheroidal galaxies sinks down rapidly. Dunlop et al. (2004)
argue that the amplification distribution derived by Perrotta et
al. (2003) is probably low because it neglects the effect of the
halo substructure. If so, the fraction of strongly lensed sources
in sub-mm surveys is even higher than shown in
Fig.~\ref{fig:counts}.

Clearly, large area sub-mm surveys can provide far richer samples
of strongly lensed sources than currently available. However, {\sc
Planck}/HFI is confusion limited at bright flux levels. Estimates
of the $5\sigma$ detection limits have been obtained by
L{\'o}pez-Caniego et al. (2006), using up-to-date templates of the
microwave/sub-mm sky, but ignoring source clustering. The effect
of latter on the confusion noise has been discussed by Negrello et
al. (2004), using models fitting the earlier $850\,\mu$m counts.
We have revised downwards the estimates by Negrello et al. (2004)
based on the counts shown in Fig.~\ref{fig:counts}. Combining
these results with those by L{\'o}pez-Caniego et al. (2006) we
obtain 5$\sigma$ detection limits of $\simeq 335$, 740, and 1290
mJy at 850, 550, and $350\,\mu$m, respectively. Note that the
analysis by L{\'o}pez-Caniego et al. (2006) refers to high
Galactic latitude regions ($|b| > 30^\circ$); at lower latitudes
the detection limits increase because of the larger contributions
of Galactic emissions to foreground fluctuations. Therefore, {\sc
Planck}/HFI can possibly pick up only the most extreme (very rare)
high-$z$ sources.

The situation is far better for {\sc Herschel}/SPIRE. For example,
the surface density of strongly lensed galaxies at a flux limit of
100 mJy is estimated to be $\simeq 0.14\,\hbox{deg}^{-2}$ at
$500\,\mu$m. Thus a survey of, say, $400\,\hbox{deg}^2$ to this
flux limit will provide a {\it complete} sample of $\simeq 60$
strongly lensed galaxies. The SASSy will provide information on
strongly lensed sources several times brighter (in terms of
bolometric luminosity) and much rarer (surface density $\sim
0.002\,\hbox{deg}^{-2}$), which however will be detected in a
comparable number ($\simeq 40$) thanks to the much larger area
($20,000\,\hbox{deg}^2$). For comparison, the radio Cosmic Lens
All-Sky Survey (CLASS; Browne et al. 2003; Myers et al. 2003),
that has been the basis for most analyses of lens statistics, has
yielded a sample of 22 gravitational lens systems.

The other source populations contributing to the counts are easily
distinguished from (lensed and unlensed) proto-spheroidal
galaxies. Essentially all radio sources are flat-spectrum blazars,
already detected by low-frequency surveys, like the 1.4 GHz NVSS
(Condon et al. 1998) and FIRST (Becker et al. 1995; White et al.
1997), complete to a few mJy levels, the 0.84 GHz SUMSS (Mauch et
al. 2003), with completeness limits ranging from 8 to 18 mJy, and
the 5 GHz GB6 survey (Gregory et al. 1986), complete to 18 mJy,
and PMN survey (Griffith \& Wright 2003) with completeness limits
ranging from 20 to 72 mJy. Altogether these surveys cover the full
sky. The median spectral index of blazars between 5 and 18.5 GHz
is (Ricci et al. 2006) $\alpha\simeq 0.16$ ($S_\nu \propto
\nu^{-\alpha}$) and generally further steepens at higher
frequencies. Therefore the radio flux of blazars is significantly
higher than at $500\,\mu$m, and thus well above the completeness
limits of the quoted surveys; obviously, the low frequency fluxes
of steep-spectrum sources are even higher.

The starburst and normal late-type galaxies sampled at the
relatively bright flux limits considered here are at $z\ll 1$ (see
Fig.~\ref{sbspzdistr}). For the median local ratios $S_{60\mu{\rm
m}}/S_{250\mu{\rm m}}\simeq 1$, $S_{60\mu{\rm m}}/S_{350\mu{\rm
m}}\simeq 2.4$, $S_{60\mu{\rm m}}/S_{550\mu{\rm m}}\simeq 9$, and
$S_{60\mu{\rm m}}/S_{850\mu{\rm m}}\simeq 41$ (Serjeant \&
Harrison 2005, based upon observational data by Dunne \& Eales
2001) these sources have $60\,\mu$m fluxes above the completeness
limit of the IRAS survey, $S_{60\mu{\rm m}}\simeq 0.6\,$Jy, for
$\lambda \ge 500\mu{\rm m}$, and well above that of the AKARI
(formerly ASTRO-F) all-sky survey, $S_{60\mu{\rm m}}\simeq
44\,$mJy (Pearson et al. 2004), at all wavelengths. Combining
SPIRE and/or SCUBA-2 with IRAS and AKARI data, photometric
redshift estimates can be obtained. On the contrary, lensed and
unlensed proto-spheroidal galaxies, at typical $z\sim 2$--3, have
$S_{60\mu{\rm m}}/S_{550\mu{\rm m}}\ll 1$.

Also, starburst and normal late-type galaxies are expected to have
optical counterparts in the available digitized sky surveys such
as SuperCOSMOS, APM, APS, DSS, and PMM. In fact only 3.5\% of the
galaxies the IRAS PSC$z$ sample (Saunders et al. 2000), complete
to $S_{60\mu{\rm m}} = 0.6\,$mJy over 84\% of the sky had no
identifications with $b_J \le 19.5\,$mag. On the other hand,
star-forming proto-spheroidal galaxies are expected to be
optically very faint ($R > 25$) because of their substantial
redshifts bringing optical bands to the rest-frame UV where
obscuration is very high.

At 500 and $850\,\mu$m, unlensed proto-spheroids have essentially
disappeared above 100 mJy so that, after having removed radio
sources and low-$z$ galaxies as discussed above, we are left with
a sample almost exclusively made of strongly lensed sources, i.e.
we expect an efficiency in the selection of such sources close to
100\%. For comparison, the CLASS survey found 22 gravitational
lens systems out of over 16000 radio sources.

The surface density of strongly lensed sources brighter than 100
mJy in the two other {\sc Herschel}/SPIRE channels is higher than
at $500\,\mu$m ($\simeq 0.6\,\hbox{deg}^{-2}$ at both $250\,\mu$m
and at $350\,\mu$m, corresponding to $\simeq 240$ strongly lensed
galaxies over $400\,\hbox{deg}^{2}$). It is however lower than
that of unlensed proto-spheroids, so that singling out strongly
lensed sources is not straightforward, although at $350\,\mu$m the
ratio of lensed to unlensed sources is still high ($\simeq 0.4$).
Also, only a fraction of starburst/late type galaxies selected at
these wavelengths are expected to show up in the IRAS catalog. On
the other hand, the different selection wavelengths provide
complementary information, since, as illustrated by
Fig.~\ref{zdistr}, they probe different redshift intervals,
systematically shifted to higher values with increasing
wavelength.

The SCUBA-2 ``Wide area extragalactic survey"  should also detect
hundreds of strongly lensed sources. In this case however, they
are only $\simeq 2\%$ of the total number of high-$z$ galaxies,
and detailed follow up observations would be necessary to single
them out. {\sc Herschel}/SPIRE surveys going substantially deeper
than 100 mJy will face a similar problem.

Clearly, the angular resolution of both {\sc Herschel}/SPIRE and
SCUBA-2 is far too poor to resolve the multiple images of lensed
sources, so that optical/IR follow-up is necessary to see the
multiple images. Note that the sources considered here are
brighter, even by large factors, than the two highly magnified
sub-mm sources discovered by Kneib et al. (2004) and Borys et al.
(2004), so that the follow-up measurements will be correspondingly
easier.

\begin{figure}
\includegraphics[width=0.95\columnwidth]{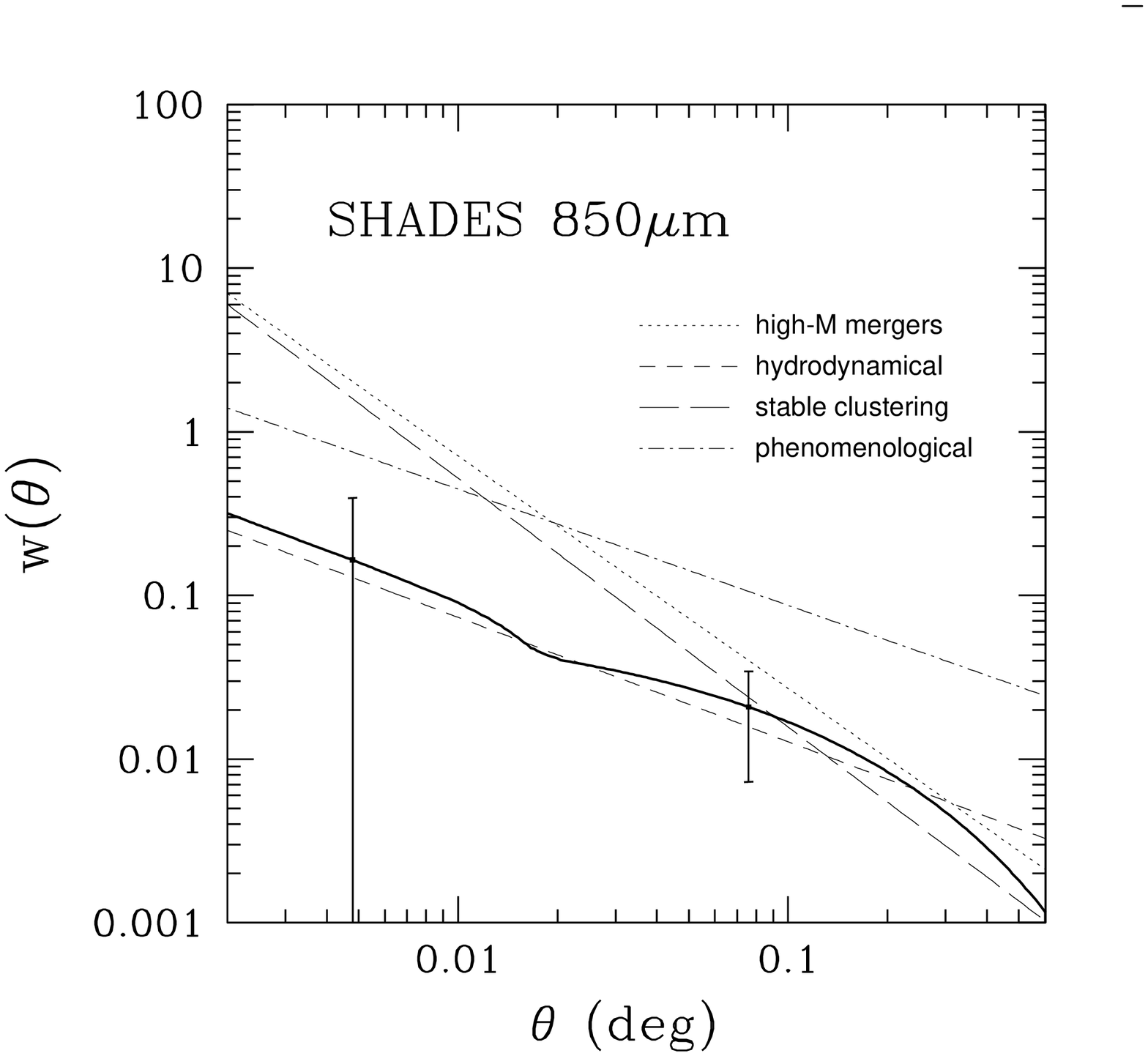}
\caption{Comparison of the angular correlation function of SHADES
sources predicted by the model adopted in this paper (solid line
with $1\sigma$ error bars) with the best fitting power laws of the
sky-averaged correlation functions predicted by the 4 models
studied by van Kampen et al. (2005).} \label{wSHADES}
\end{figure}

\section{Clustering of proto-spheroidal galaxies}

Both theoretical arguments and observational data indicate that
the positions of powerful far-IR galaxies, detected by SCUBA
surveys and interpreted as massive proto-spheroidal galaxies, are
highly correlated (see e.g. Smail et al. 2003; Negrello et al.
2004; Blain et al. 2004; Scott et al. 2006). On the contrary,
Poisson fluctuations dominate in the case of the other
extragalactic source populations contributing to the sub-mm
counts, whose clustering is relatively weak, as in the case of
spiral and starburst galaxies (cf. e.g. Madgwick et al. 2003) or
highly diluted because of the very broad redshift distribution, as
in the case of radio sources (Magliocchetti et al. 1999; Blake \&
Wall 2002a,b; Overzier et al. 2003; Blake et al. 2004; Negrello et
al. 2006).

In the following we will therefore consider only the clustering of
dusty proto-spheroidal galaxies. As already mentioned, they are
located at substantial ($\gsim 2$) redshifts (Chapman et al. 2005)
and display huge star formation rates ($\gsim
1000\,$M$_{\odot}$/yr), allowing masses in stars of the order of
10$^{11}$ M$_{\odot}$ to be assembled in times shorter than 1 Gyr.

For dark matter to stellar mass ratios typical of massive
ellipticals (see, e.g., Marinoni \& Hudson 2002; McKay et al.
2002), the dark matter haloes in which the sub-millimeter galaxies
reside have masses of $\geq 10^{13}$ M$_{\odot}$. Since such
massive haloes sample the rare high-density peaks of the
primordial dark matter distribution (Kaiser 1984; Mo \& White
1996), these sources are expected to exhibit a strong spatial
clustering, similar to that measured for Extremely Red Objects
(EROs; Daddi et al. 2001, 2003), for which similar masses have
been inferred (Moustakas \& Somerville 2002).

Direct measurements of clustering properties of SCUBA galaxies are
made difficult by the poor statistics and by the fact that they
are spread over a wide redshift range, so that their clustering
signal is strongly diluted. However, tentative evidences of strong
clustering with a {\it comoving} correlation length $r_0 \sim
8$--13$\,\hbox{h}^{-1}$ Mpc, consistent with that of EROs, have
been reported (Scott et al. 2002, 2006; Webb et al. 2003; Smail et
al. 2003; Blain et al. 2005). A highly statistically significant
detection of strong clustering for ultraluminous infrared galaxies
over $1.5 < z < 3$ has been obtained by Farrah et al. (2006a,b)
from an analysis of optically faint sources selected at $24\mu$m
in three fields observed as a part of the {\it Spitzer} Wide-area
Infrared Extragalactic (SWIRE) survey. They found $r_0 = 14.40\pm
1.99 \,\hbox{h}^{-1}$ Mpc for $2 < z <3$ and $r_0 = 9.40\pm 2.24
\,\hbox{h}^{-1}$ Mpc for $1.5 < z <2$. Magliocchetti et al. (2006)
found that optically unseen ($R>25.5$) $24\mu$m selected sources
in the complete {\it Spitzer} First Look Survey sample (Fadda et
al. 2006) are highly clustered and pointed out several lines of
circumstantial evidence consistently indicating that they are at
$z \sim 2$. If so, their comoving clustering length is $r_0 =
10.6^{+1.6}_{-1.8}\,\hbox{h}^{-1}$ Mpc.

\begin{figure*}
\includegraphics[width=0.90\textwidth]{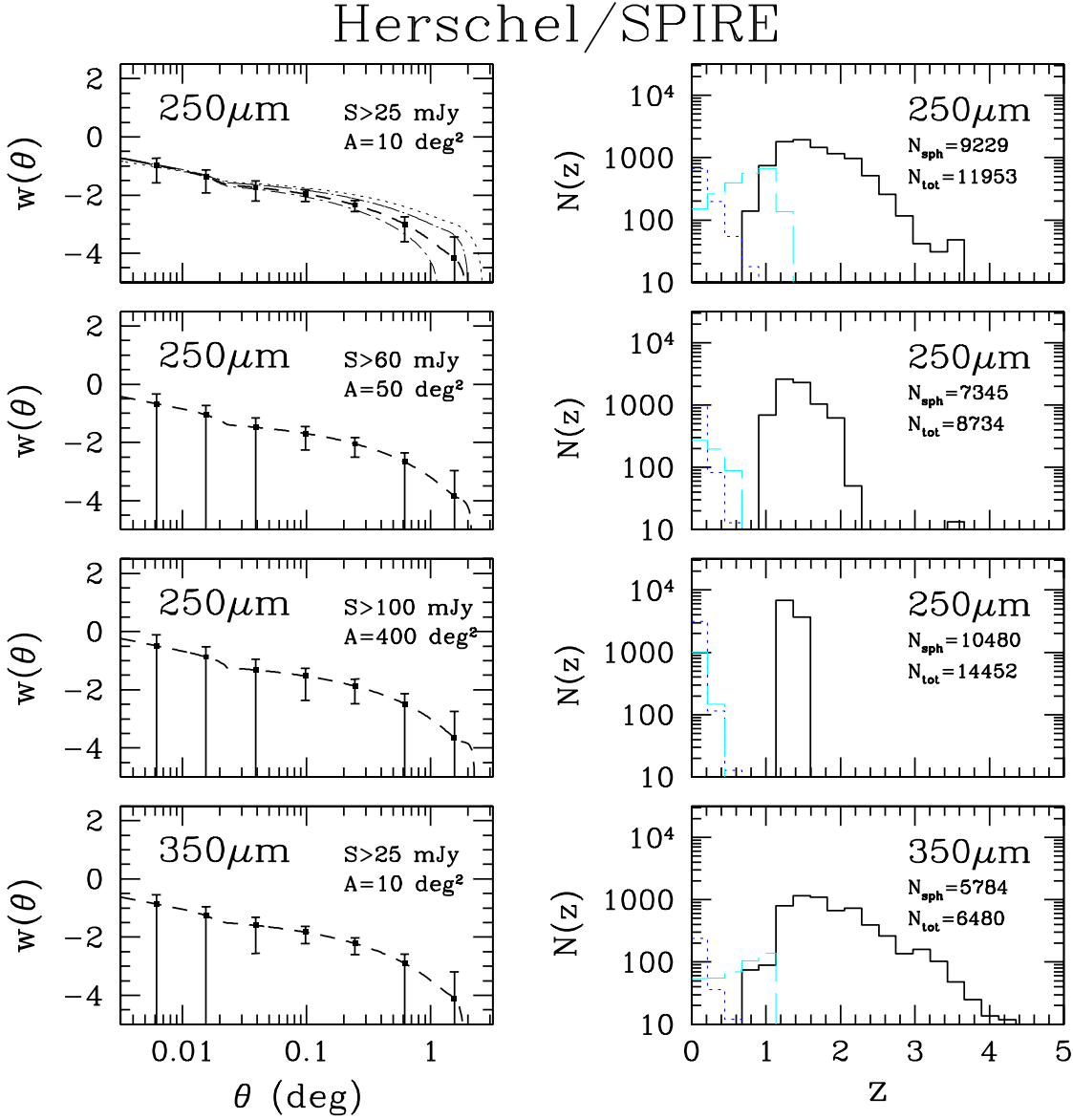}
\caption{Predicted angular correlation function, with its
uncertainties ($3\sigma$ Poisson error bars), of proto-spheroids
detected by $250\,\mu$m surveys to flux limits of 25, 60, and 100
mJy, over areas of 10, 50, and 100$\,\hbox{deg}^2$, respectively,
and by a $350\,\mu$m survey of 10$\,\hbox{deg}^2$ to a flux limit
of 25 mJy, averaged over bins $\Delta\log(\theta)=0.4$. The
dilution of the signal due to the other, weakly clustered, source
populations is taken into account. The corresponding predicted
numbers of detected sources and their redshift distributions for
redshift bins of $\Delta z = 0.23$ are also shown. The predicted
redshift distributions of spiral and starburst galaxies (dotted
and dashed histograms, respectively) are also shown. Surveys at
$350\,\mu$m to flux limits of 60 and 100 mJy over areas of 50 and
400$\,\hbox{deg}^2$ would provide only very weak constraints on
$w(\theta)$. Deep surveys, reaching the estimated confusion limits
at 250 and $350\,\mu$m, can provide useful constraints on
cosmological parameters, as illustrated by the top left-hand
panels where, in addition to the predicted $w(\theta)$ for our
reference cosmological model, with $\Omega_m=0.3$ and
$\Omega_b=0.047$ (dashed line), we show the effect of different
choices of these 2 parameters: $\Omega_m=0.15$ and
$\Omega_b=0.045$ (dotted line); $\Omega_m=0.27$ and $\Omega_b=0.$
(dot-short dashed line); $\Omega_m=0.27$ and $\Omega_b=0.09$
(dot-long dashed line).
 }\label{wthdNdz}
\end{figure*}

\subsection{Formalism}\label{sec:form}

We model the two-point spatial correlation function, $\xi(r,z)$,
of star-forming spheroids as in Negrello et al. (2004; their model
1):
\begin{equation}
\xi(r,z)=b^{2}(M_{\rm eff},z)\xi_{\rm DM}(r,z), \label{eq:xi_lin}
\end{equation}
where $b(M_{\rm eff},z)$ is the redshift-dependent (linear) bias
factor (Sheth $\&$ Tormen 1999; Percival et al. 2003), $M_{\rm
eff}$ being the effective mass of the dark matter halos in which
the sources reside; $\xi_{\rm DM}$ is the linear two-point spatial
correlation function of dark matter, determined by the power
spectrum of primordial density perturbations as well as by the
underlying cosmology. For the power spectrum of the primordial
fluctuations we adopt the fitting relations by Eisenstein \& Hu
(1998) which account for the effect of baryons on the matter
transfer function.

We adopt an effective halo mass of star-forming spheroids $M_{\rm
eff}=10^{13}\hbox{h}^{-1}$ M$_{\odot}$, which yields a comoving
clustering length, defined by $\xi(r_0,z)=1$, $r_0 \simeq
8\,\hbox{h}^{-1}$ Mpc, consistent with the available observational
indications for SCUBA galaxies (see Negrello et al. 2004), and
somewhat below the estimates for {\it Spitzer} $24\,\mu$m sources
at $z\simeq 2$. Note that, on scales below a few Mpc, clustering
enters the non-linear regime and the slope of $\xi(r,z)$ becomes
steeper than expected from the linear theory. In this range of
scales, a power-law model provides a better description of the
two-point spatial correlation function. Therefore we assume
\begin{equation}
\xi(r,z)=[r/r_{0}(z)]^{-1.8}\ \hbox{for}\ r<2r_{\rm vir}(M_{\rm
eff},z)\sim 2\,\hbox{Mpc}, \label{eq:xi_nolin}
\end{equation}
$r_{\rm vir}$ being the comoving virial radius of the
characteristic halo in which star-forming spheroids reside. In
view of the tight connection between spheroidal galaxies and
active galactic nuclei at their centers entailed by the model of
Granato et al. (2004), we assume the correlation length, $r_{0}$,
to be constant in comoving coordinates, as suggested by quasar
data (Porciani et al. 2004; Croom et al. 2005).

The angular correlation function is related to the spatial one,
$\xi(r,z)$, by the relativistic Limber equation (Peebles, 1980):
\begin{eqnarray}
w(\theta)=2\:\frac{\int_0^{\infty}\int_0^{\infty}F^{-2}(x)x^4\Phi^2(x)
\xi(r,z)dx\:du}{\left[\int_0^{\infty}F^{-1}(x)x^2\Phi(x)dx\right]^2},
\label{eqn:limber}
\end {eqnarray}
where $x$ is the comoving radial coordinate, $u$ is defined by
$r=\{u^2+[x\theta/(1+z)]^2\}^{1/2}$, $F(x)$ gives the correction
for curvature, and the selection function $\Phi(x)$ satisfies the
relation
\begin{eqnarray}
{\cal N}=\int_0^{\infty}\Phi(x) F^{-1}(x)x^2 dx=\frac{1}{\Omega_s}
\int_0^{\infty}N(z)dz, \label{eqn:Ndense}
\end{eqnarray}
in which $\cal N$ is the mean surface density, $\Omega_s$ is the
solid angle covered by the survey, and $N(z)$ is the number of
sources within the shell ($z,z+dz$), given by the model.


The errors on $w(\theta)$ have been computed as:
\begin{eqnarray}
\delta w(\theta) = {1+ w(\theta) \over \sqrt{N_{\rm pair}}}\
,\label{eq:errw}
\end{eqnarray}
where $N_{\rm pair}$ is the number of source pairs separated by an
angular distance $\theta$.

\begin{figure*}
\includegraphics[width=0.95\textwidth]
{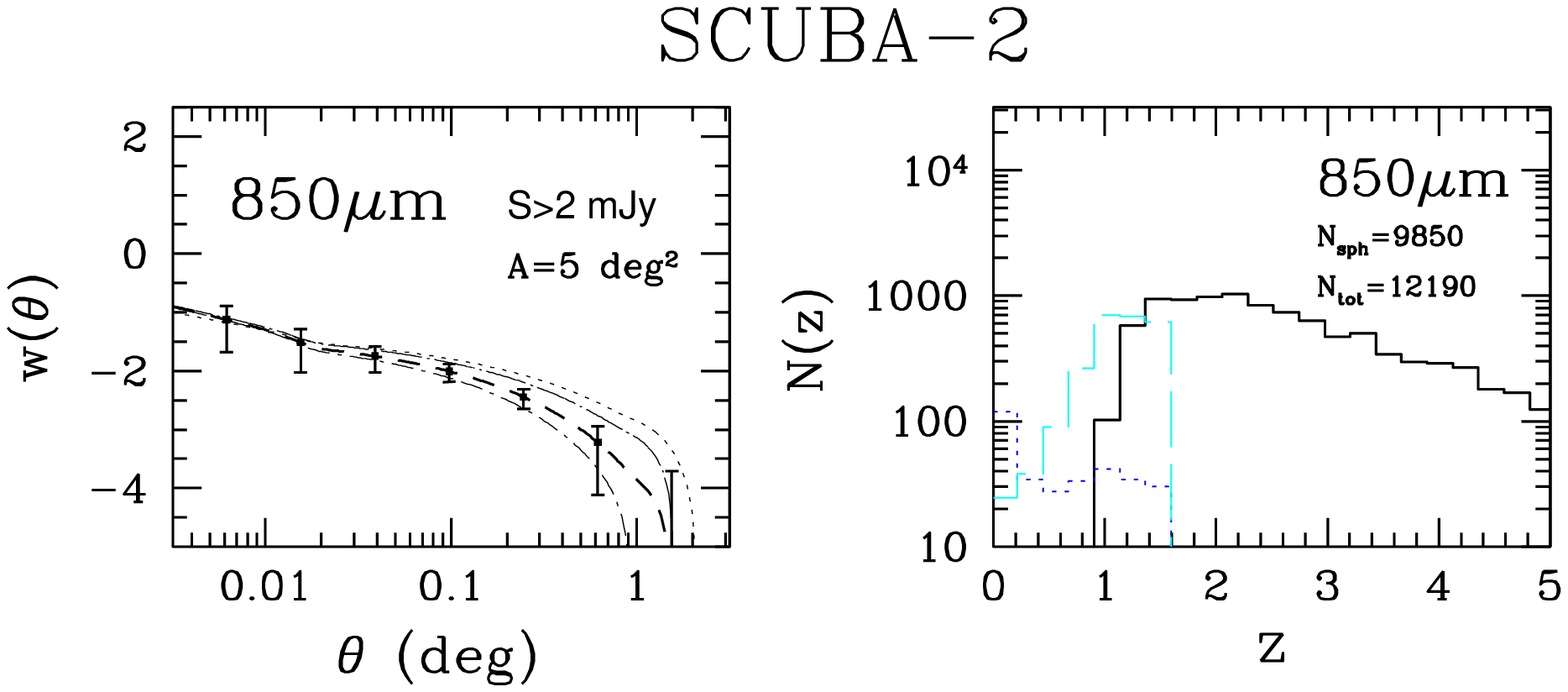} \caption{Predicted angular
correlation function, with its uncertainties ($3\sigma$ Poisson
error bars), of proto-spheroids detected by the $850\,\mu$m
SCUBA-2 survey of 2$\,\hbox{deg}^2$ to a detection limit of 2 mJy,
and the corresponding redshift distribution. The dilution of the
signal due to the other, weakly clustered, source populations is
taken into account. The predicted redshift distributions of spiral
and starburst galaxies (dotted and dashed histograms,
respectively) are also shown. We have used bins of
$\Delta\log(\theta) = 0.2$ and $\Delta z = 0.23$. The different
lines in the left-hand panel correspond to the same choices of the
cosmological parameters $\Omega_m$ and $\Omega_b$ as in
Fig.~{\protect\ref{wthdNdz}}. } \label{wthdNdzSCUBA2}
\end{figure*}

The angular correlation function of intensity fluctuations due to
sources below the detection limit writes (Peebles 1980; De Zotti
et al. 1996)
\begin{eqnarray}
C_C(\theta)&=&\left({1\over 4\pi}\right)^2  \int_{z_{(L_{\rm
min},S_d})}^{z_{\rm max}}\!\!\!\!\!\!\!\!\! dz\;b_{\rm
eff}^2(M_{\rm eff},z)\;
\frac{j^2_{\rm eff}(z)} {(1+z)^4}\left(\frac{dx}{dz}\right)^2\nonumber\\
&\cdot& \int_0^\infty d(\delta z)\;\xi(r,z), \label{eq:cth}
\end{eqnarray}
$z_{\rm max}$ being the redshift when the sources begin to shine,
$z(S_d,L)$ the redshift at which a source of luminosity $L$ is
seen with a flux equal to the detection limit $S_d$, and $\delta
z$ the redshift difference of sources separated by a comoving
distance $r$. The effective volume emissivity $j_{\rm eff}$ is
expressed as:
\begin{eqnarray}
j_{\rm eff}=\int_{L_{\rm min}}^{{\rm min}[L_{{\rm
max},L(S_d,z)}]}\!\!\!\!\!  \Phi(L,z)\; K(L,z)\; L\;d{\rm log}L,
\label{eq:jeff}
\end{eqnarray}
$\Phi(L,z)$ being the luminosity function (per unit logarithmic
luminosity interval), $K(L,z)$ the K-correction and $L_{\rm max}$
and $L_{\rm min}$ the maximum and minimum local luminosity of the
sources.

The angular power spectrum of the intensity fluctuations can then
be obtained as
\begin{eqnarray}
C_\ell=\langle |a_l^0|^2 \rangle=\int_0^{2\pi}\int_0^\pi
C_C(\theta)\; P_l(\cos\theta)\;\sin\theta\;d\theta\;d\phi\ .
\end{eqnarray}
%
%
The fractional error on the power spectrum can be estimated
summing in quadrature the instrumental, confusion and cosmic
variances (Knox 1995):
\begin{equation}
{\delta C_\ell \over C_\ell} = \left({4\pi \over A}\right)^{1/2}
\left({2\over 2\ell +1}\right)^{1/2} \left(1 + {A\sigma^2 \over N
C_\ell W_\ell}\right)\ , \label{knox}
\end{equation}
where $A$ (sr) is the surveyed area, $\sigma$ is the rms noise per
pixel (quadratic sum of instrumental and confusion noise), $N$ is
the number of pixels in the map, and $W_\ell$ is the window
function, describing the region of the $\ell$-space observable
with the considered instrument. We assume Gaussian beams with Full
Width at Half Maximum $\hbox{FWHM}=2\sigma_B\sqrt{2\ln 2}$, so
that $W_\ell=\exp(-\ell^2\sigma_B^2)$.

In the dependence on the angular scale of both the angular
correlation functions and the power spectra of background
fluctuations due to clustering, three regimes can be distinguished
(see Cooray \& Sheth 2002 for a review). On small scales the power
spectrum probes the statistics of the distribution of sources
(sub-halos) within large-scale dark matter halos i.e. the halo
occupation function, determined by the non-linear evolution of
large-scale structure. On intermediate scales, the amplitude of
the signal is mostly determined by the bias parameter, which in
turn depends on the effective mass of dark halos associated to
sources. On very large scales, the signal provides information on
the cosmological evolution of primordial density fluctuations in
the linear phase, and is thus sensitive to cosmological
parameters, and in particular to the dark matter and baryon
densities.

\begin{figure*}
\includegraphics[width=0.90\textwidth]{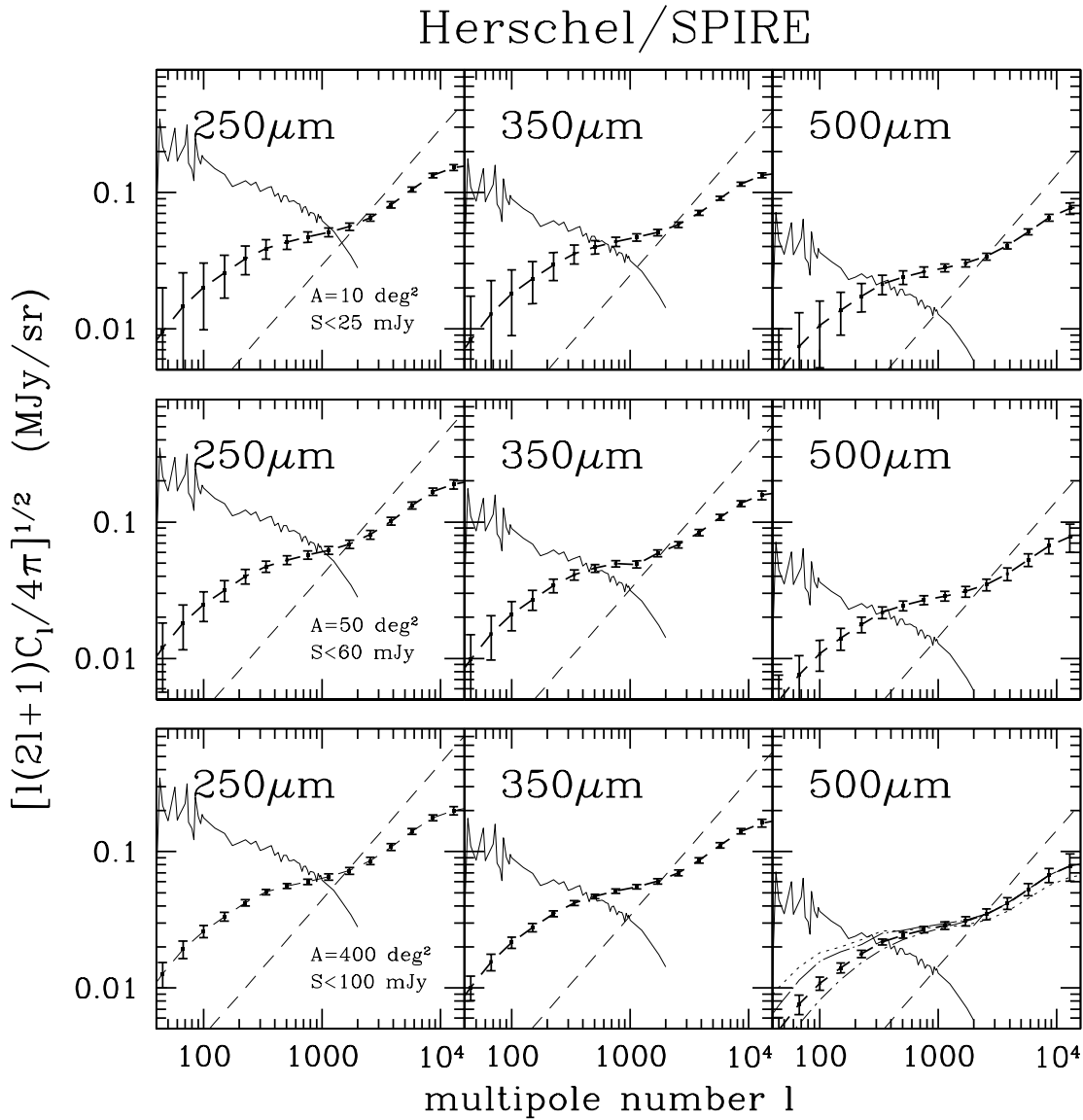}
\caption{Power spectra of background fluctuations due to
clustering of proto-spheroidal galaxies below the detection limit
compared with the power spectra of the diffuse Galactic dust
emission and of Poisson fluctuations. The ``data'' points refer,
from top to bottom, to {\sc Herschel}/SPIRE surveys of 10, 50, and
$400\,\hbox{deg}^2$ with detection limits of 25, 60, and 100 mJy,
respectively, for all channels (250, 350, and $550\,\mu$m). The
error bars are estimated from eq.~(\protect\ref{knox}), averaging
over bins of $\Delta \ell = 0.4\ell$.  The broken lines show the
power spectra of Galactic dust emission based on the Finkbeiner et
al. (1999) model, averaged over  a particularly dust-free
$400\,\hbox{deg}^2$ area centered at $l\simeq 245^\circ$, $b\simeq
-50^\circ$.
The dashed lines show the Poisson contributions of extragalactic
point sources below the detection limits. In the bottom right-hand
panels we show, in addition to the predicted power spectrum for
our reference cosmological model, with $\Omega_m=0.3$ and
$\Omega_b=0.047$ (dashed line), the effect of different choices of
these 2 parameters: $\Omega_m=0.15$ and $\Omega_b=0.045$ (dotted
line); $\Omega_m=0.27$ and $\Omega_b=0.$ (dot-short dashed line);
$\Omega_m=0.27$ and $\Omega_b=0.09$ (dot-long dashed line). }
\label{powerspSpire}
\end{figure*}

\subsection{Predictions for $w(\theta)$ and $C_\ell$}

Before proceeding to describe our predictions it is interesting to
compare our model with the four alternative models for the clustering
of SCUBA galaxies investigated by van Kampen et al.  (2005). These
include a simple merger model, a hydrodynamical model, the stable
clustering model by Hughes \& Gazta\~naga (2000), and a
phenomenological model. For each of these models van Kampen et
al. (2005) worked out estimates of the clustering properties, tailored
for the SHADES survey. In Fig.\,\ref{wSHADES} these estimates are
compared with those of the model used here.  There are clear
differences among the models, so that the analysis of clustering will
help to discriminate between them. The $w(\theta)$ predicted by our
model is remarkably close to that of the hydrodynamical model, which
assumes a redshift distribution (the analytical form of Baugh et
al. 1996) close to being compatible with the results by Aretxaga et
al. (2007).

\begin{figure*}
\includegraphics[width=0.90\textwidth]{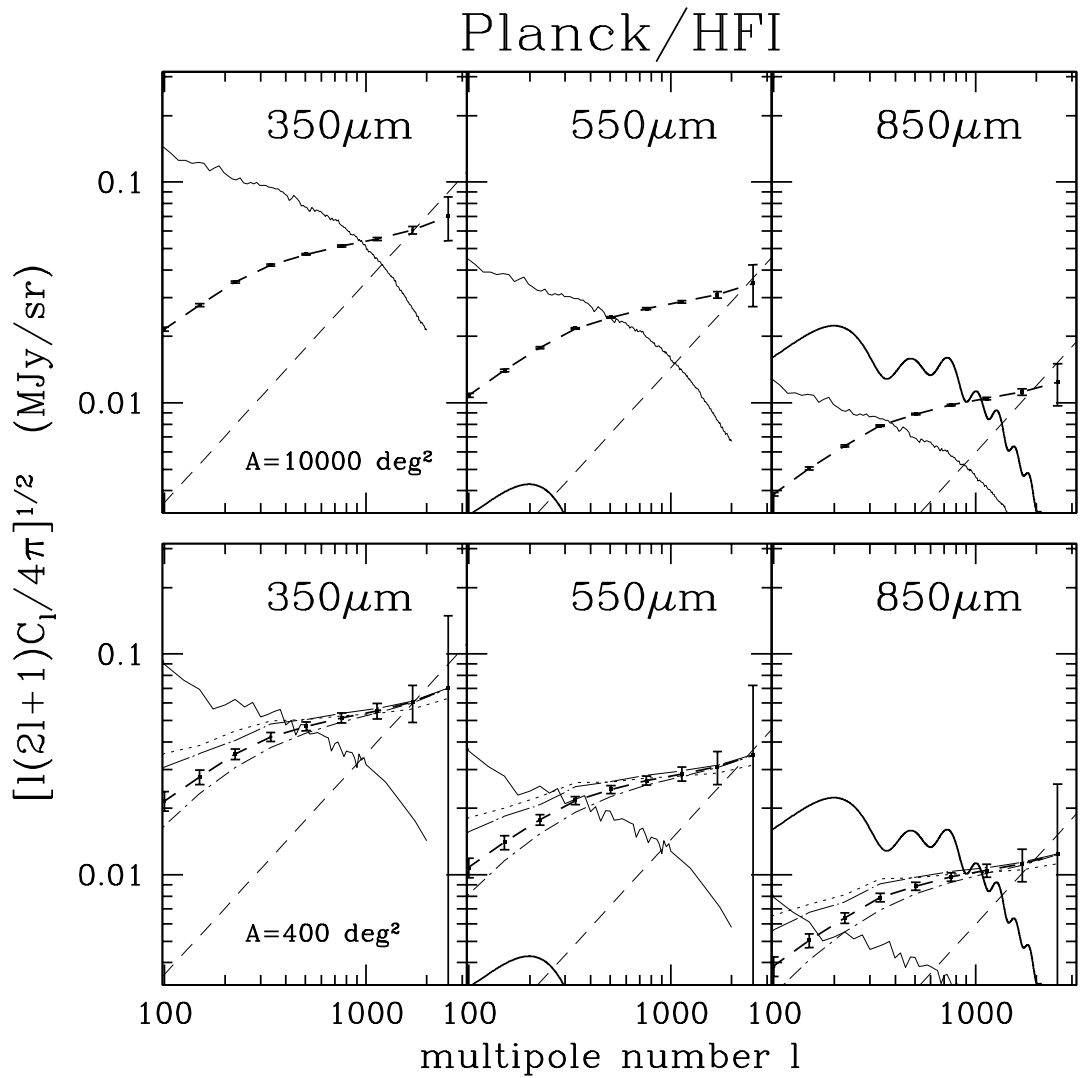}
\caption{Power spectra of background fluctuations due to
clustering of proto-spheroidal galaxies below the detection limit
in the high frequency {\sc Planck}/HFI surveys (points with error
bars for bins $\Delta \ell = 0.4\ell$) compared with the power
spectra of the diffuse Galactic dust emission (broken lines), of
Poisson fluctuations due to extragalactic point sources below the
detection limits (dashed lines), and of the CMB (solid line,
visible in the $850\,\mu$m panel, and in the bottom left-hand
corner of the $550\,\mu$m panel). The dust power spectra are based
on the Finkbeiner et al. (1999) model; the upper set of panels
refer to the $\simeq 9650\,\hbox{deg}^2$ at $|b|>50^\circ$, the
lower set to the $400\,\hbox{deg}^2$ low-dust region centered at
$l\simeq 245^\circ$, $b\simeq -50^\circ$.
In the bottom panels we show, in addition to the predicted power
spectrum for our reference cosmological model, with $\Omega_m=0.3$
and $\Omega_b=0.047$ (dashed line), the effect of different
choices of these 2 parameters: $\Omega_m=0.15$ and
$\Omega_b=0.045$ (dotted line); $\Omega_m=0.27$ and $\Omega_b=0.$
(dot-short dashed line); $\Omega_m=0.27$ and $\Omega_b=0.09$
(dot-long dashed line).} \label{powerspHFI}
\end{figure*}

Our predictions for the angular correlation function, $w(\theta)$,
and/or the power spectrum of intensity fluctuations, $C_\ell$, for
the large area surveys defined in the \S\,1 are presented in
Figs.~\ref{wthdNdz}--\ref{powerspHFI}.






The steep counts of proto-spheroids shown by Fig.~\ref{fig:counts}
imply that in the case of {\sc Planck}/HFI and of an {\sc
Herschel}/SPIRE survey flux limited at 100 mJy at $500\,\mu$m the
clustering signal comes entirely from sources below the detection
limit. For the same flux limit, we expect significant signals both
from sources above and below threshold at 350 and $250\,\mu$m.
Only at the latter wavelength, however, a 400$\,\hbox{deg}^2$
survey would provide a large enough sample of detected spheroidal
galaxies allowing a direct study of their angular correlation
function (see Fig.~\ref{wthdNdz}). A deeper survey (flux limit of
$\simeq 25\,$mJy) on one hand provides a substantially larger
sample in a given observing time (thanks to the steepness of the
counts of proto-spheroids in this flux density interval) but, on
the other hand, contains a higher fraction of weakly clustered
late-type (starburst and spiral), diluting the signal. The former
effect, however, prevails, so that, as shown by
Fig.~\ref{wthdNdz}, we expect a rather accurate determination of
$w(\theta)$ particularly at $250\,\mu$m, where the data will
effectively probe all the three clustering regimes mentioned
above. The number of detected sources decreases rapidly, for given
area and limiting flux, with increasing wavelength. Already at
$350\,\mu$m the uncertainties are rather large. On the other hand,
studies at longer wavelengths explore higher $z$'s (because of the
effect of the strongly negative K-correction).

The SASSy survey will detect, apart from the strongly lensed
sources discussed in \S\,\ref{sec:lensing}, mostly blazars and
low-$z$ late-type/starburst galaxies. The size of the samples is
unlikely to be sufficient to detect the weak clustering of these
populations. On the other hand, the SCUBA-2 `Wide-area
extragalactic Survey' is well suited for the investigation of
small- (non-linear), intermediate (sensitive to the bias
parameter), and large-scale (sensitive to cosmological parameters)
clustering, since it has a very good statistics and the clustering
signal is less diluted than in the (already very fit) deepest
$250\,\mu$m and $350\,\mu$m {\sc Herschel}/SPIRE surveys
considered in Fig.~\ref{wthdNdz}. It also covers a much wider
redshift range (see Fig.~\ref{wthdNdzSCUBA2}).

{\sc Herschel}/SPIRE surveys at $500\,\mu$m, which are expected
not to provide large enough samples of individually detected
high-$z$ proto-spheroids to allow accurate estimates of their
$w(\theta)$, are however better suited than surveys at shorter
wavelengths to estimate the autocorrelation function of background
fluctuations, from which the clustering power spectra can be
derived. This is illustrated by Fig.~\ref{powerspSpire} where the
clustering signal is compared with competing astrophysical
sources, namely the Poisson fluctuations due to the combination of
all populations of extragalactic sources fainter than the
detection limit, and to Galactic dust emission. To deal with the
latter we have used the Finkbeiner et al. (1999) model, built
combining IRAS and COBE data. We have assumed that the {\sc
Herschel} survey is carried out in the low-dust region centered at
$l\simeq 245^\circ$, $b\simeq -50^\circ$. Because of the effect of
redshift, the clustering power spectrum of high-$z$
proto-spheroids dominates over that of dust (in the chosen region)
over a multipole range increasing with increasing wavelength. At
$500\,\mu$m, the `clean' range is broad enough to provide
information both on the bias factor and on cosmological parameters
in each of the three considered cases, but the highest accuracy is
achieved by the largest area, shallow survey. On the contrary, the
possibility of taking full advantage of the {\sc Herschel}/SPIRE
resolution to learn about the non-linear clustering regime (whose
onset is marked by the high-multipole inflection point) is
seriously constrained by Poisson fluctuations.

While the much larger area covered by {\sc Planck} allows, in
principle, a much more accurate estimate of the power spectrum of
the clustering signal, this advantage is largely impaired by the
contamination by Galactic dust emission. For example, even if we
restrict ourselves to the $\simeq 9650\,\hbox{deg}^2$ around the
Galactic polar regions ($|b| > 50^\circ$), the interstellar dust
emission power spectrum blurs the clustering contribution for
multipoles $\ell \lsim 500$ at $550\,\mu$m and for $\ell \lsim
900$ at $350\,\mu$m (Fig.~\ref{powerspHFI}). The effect of
Galactic dust decreases substantially at $850\,\mu$m, but then the
power spectrum is dominated by CMB anisotropies up to $\ell \simeq
1000$. On the other hand, it is possible to select regions of
$\sim 400\,\hbox{deg}^2$ where the Galactic dust contamination is
much lower and below the clustering signal down to $\ell \lsim
350$ at $550\,\mu$m and to $\ell \lsim 500$ at $350\,\mu$m. Again,
at high multipoles ($\ell \gsim 2000$), Poisson fluctuations
dominate. An application of eq.~(\ref{knox}) shows that the
uncertainties on the clustering power spectrum on scales much
larger that the instrument FWHM are, at given wavelength, almost
identical for {\sc Planck} and {\sc Herschel} surveys of the same
area. The substantial increase of the error bars at $\ell \simeq
2500$ signals that we are approaching the limit set by {\sc
Planck}'s resolution.

\section{Other source populations}\label{blazar}

\subsection{Local late-type and starburst galaxies}

According to the counts shown in Fig.~\ref{fig:counts}, an
Herschel/SPIRE survey of $400\,\hbox{deg}^2$ to a flux limit of
$100\,$mJy will yield $\simeq 3000$ low-$z$ galaxies at
$250\,\mu$m, $\simeq 890$ at $350\,\mu$m, and $\simeq 190$ at
$500\,\mu$m. Low-$z$ galaxies are expected to dominate the counts
at these bright flux levels, except perhaps at $250\,\mu$m. In any
case, they can be easily distinguished from the other populations
as described in \S\,3. The SCUBA-2 SASSy survey will yield about
350 low-$z$ galaxies at $850\,\mu$m. Adopting the $5\sigma$
detection limits estimated in \S\,3, we find that number of
low-$z$ galaxies detected by {\sc Planck}/HFI in the 50\% of the
sky at $|b|>30^\circ$ will be of 110, 300, and 990 at $850\,\mu$m,
$550\,\mu$m, and $350\,\mu$m, respectively.

Thus, these surveys will provide large enough samples of low-$z$
galaxies to allow the first {\it direct} and accurate estimates of
the local luminosity functions of late-type and starburst galaxies
at several sub-mm wavelengths.

\subsection{Radio sources}

Based on the current understanding of the blazar spectra, the
slope, $\beta$, of the integral counts ($N(>S)\propto S^{-\beta}$)
of blazars in the relevant flux density range cannot be much
different from that measured by high-frequency radio surveys
(Waldram et al. 2003; Ricci et al. 2004), i.e. $\beta \simeq 1$.
Then, the number of detected sources in an observing time $t$
increases as $t^{1/2}$ if we use the time to go deeper over a
fixed area, and increases as $t$ if we cover a larger area to a
fixed depth. Thus, a larger area shallow survey is substantially
more productive than a deep survey over a small area. From the De
Zotti et al. (2005) model we expect that the SASSy survey detects
$\simeq 1,200$ blazars, and that the {\sc Planck} surveys detect
$\simeq 550$, 130, and 50 sources at $850\,\mu$m, $550\,\mu$m, and
$350\,\mu$m, respectively, for the flux limits given in \S\,3 and
at $|b|> 30^\circ$.  But since a key scientific goal is the
identification of the synchrotron peak frequency, a broad spectral
coverage is essential and large area surveys at {\sc
Herschel}/SPIRE wavelengths would also be extremely valuable. A
$400\,\hbox{deg}^2$ survey to a flux limit of $100\,$mJy is
expected to provide a {\it complete} sample of $\simeq 30$ blazars
detected at each SPIRE wavelength. As explained in \S\,3, the
radio source populations can be easily identified by a positional
cross correlation with radio catalogues, covering the whole sky at
at least 2 frequencies.

\section{Discussion and conclusions}\label{sec:conc}

Combining the predictions of the physical model by Granato et al.
(2004) for the formation and evolution of spheroidal galaxies
(updated to take into account the most recent $850\,\mu$m counts)
with up-to-date phenomenological models for the evolution of
starburst and normal late-type galaxies and of radio sources, we
have worked out quantitative predictions for the many important
wide-area, shallow surveys that will be carried out in the next
few years.

We find that {\sc Planck}/HFI surveys are confusion limited at
bright flux levels, with $1\,\sigma$ confusion noise increasing
from $\simeq 70\,$mJy at $850\,\mu$m to $\simeq 260\,$mJy at
$350\,\mu$m. These surveys will thus be not sufficiently deep to
detect with good signal-to-noise ratio ($S/N >5$) high-$z$
galaxies, except in the case of extreme gravitational
magnifications or exceptionally high luminosities. They will
however, provide complete samples of hundreds low-$z$ star-forming
(starburst and normal late-type) galaxies, enabling a direct,
accurate, determination of their local luminosity functions at
sub-mm wavelengths. In addition, they will detect, at least at
$850\,\mu$m, several hundreds blazars, providing important
information on their physical properties.

Although the number of sub-mm sources of each population detected
by {\sc Planck} is too small to allow a measurement of their
angular correlation function (unless their clustering is far
stronger than indicated by current data), the {\sc Planck} maps
will allow accurate determination of the autocorrelation function
of intensity fluctuations due to clustering of proto-spheroidal
galaxies. It will however be difficult to take full advantage of
{\sc Planck}'s sky coverage because of the contamination by
Galactic dust. Even at $|b| > 50^\circ$, the Galactic dust power
spectrum is expected to overwhelm that of clustering for multipole
numbers $\ell < 1000$ in the $350\,\mu$m channel and for $\ell \le
600$ in the $500\,\mu$m channel (see Fig.~\ref{powerspHFI}).
Nevertheless, the useful range of angular scales provides precise
information on the bias parameter of halos hosting the sub-mm
sources, hence on their masses.

There are however smaller low-dust regions where the clustering
power spectrum may dominate over that of Galactic dust down to
$\ell \sim 400$--500 at $550\,\mu$m and at $350\,\mu$m,
respectively. Such regions are large enough to allow us to
investigate the regime where the shape of the clustering power
spectrum is sensitive to the cosmological parameters (especially
to the baryon density and to the matter density), so that its
accurate determination sets interesting constraints on them. At
$850\,\mu$m, the dominant signal up to $\ell \simeq 1000$ is the
Cosmic Microwave Background (CMB). For high multipoles, $\ell
\gsim 2500$, {\sc Planck}'s ability to measure the power spectrum
hits the limit set by angular resolution; however, already at
somewhat lower multipoles ($\ell \sim 2000$) Poisson fluctuations
overcome the clustering ones.

One might expect that, thanks to its much higher angular
resolution (compared to {\sc Planck}/HFI) {\sc Herschel}/SPIRE can
remove much more efficiently point sources and thus substantially
decrease the Poisson fluctuation level. However, while it is
obviously true that {\sc Herschel}/SPIRE can reach much fainter
detection limits, the very steep sub-mm counts of proto-spheroidal
and starburst galaxies imply that the dominant contribution to
Poisson fluctuations come from the still fainter fluxes where the
slope of integral counts flattens below the critical value $\beta
=2$. Thus while, in principle, the {\sc Herschel}/SPIRE resolution
would allow us to reach very high multipoles (up to $\ell \sim
2\times 10^4$), the Poisson fluctuations are expected to dominate
for $\ell \gsim 2 \times 10^3$, i.e. over the multipole range
where 1-halo contributions (i.e. the clustering of sources within
a single dark matter halo) come out, even for surveys reaching the
confusion limits. On the whole, we do not expect that {\sc
Herschel}/SPIRE surveys will add much to clustering power spectrum
determinations provided by {\sc Planck}/HFI, except for the {\sc
Herschel}-only $250\,\mu$m channel. They will however provide
independent estimates.

The great advantage of {\sc Herschel}/SPIRE over {\sc Planck}/HFI
resides in its capability of detecting large enough source samples
to allow a direct determination of the angular correlation
function $w(\theta)$, at least at $250\,\mu$m and at $350\,\mu$m
(Fig.~\ref{wthdNdz}). In particular, a $250\,\mu$m survey of
$10\,\hbox{deg}^2$ reaching a flux limit of 25 mJy effectively
probes all the three clustering regimes mentioned in
\S\,\ref{sec:form}. As illustrated by Fig.\,\ref{wSHADES},
measurements of the $w(\theta)$ are also useful to discriminate
between galaxy formation models (van Kampen et al. 2005; Percival
et al. 2003), although the error bars are expected to be rather
large. Surveys of different depths probe different redshift
intervals.

At 350 and $500\,\mu$m, {\sc Herschel}/SPIRE surveys covering
$\sim 400\,\hbox{deg}^2$ are necessary to extend the counts up to
flux density levels sampled by {\sc Planck}, which, at these
wavelengths, is confusion limited at $\sim 1\,$Jy. A survey of a
similar area, to a $5\sigma$ detection limit $\simeq 100\,$mJy,
would also provide sufficient samples of $z\lsim 0.1$
late-type/starburst galaxies to estimate their local luminosity
functions.

At $500\,\mu$m, such a very large area, shallow survey will also
be ideal to select $\simeq 60$ strongly lensed high-$z$ dusty
galaxies with an essentially 100\% efficiency. As discussed in
\S\,3, the other source populations can be easily distinguished
using only available photometric data.  The simultaneous SPIRE
surveys at shorter wavelengths with a similar detection limit will
detect larger numbers of strongly lensed sources (e.g. we expect
$\simeq 240$ strongly lensed galaxies at $350\,\mu$m), but
singling them out will be a less easy and efficient process,
although at $350\,\mu$m the ratio of lensed to unlensed
proto-spheroidal galaxies is estimated to be still high ($\simeq
0.4$). The redshift distributions of lensed sources is
increasingly shifted to higher redshifts with increasing survey
wavelength.

We have also estimated that the SASSy survey will detect $\sim 40$
strongly lensed sources, substantially brighter (in terms of
bolometric luminosity) and with more extreme amplifications than
those detected by {\sc Herschel}/SPIRE surveys with a flux limit
of $\sim 100\,$mJy. These sources will be retrievable with an
essentially $100\%$ efficiency, as in the case of the latter
surveys.

On the other hand, in spite of its very large area, the SASSy
survey will probably not detect enough proto-spheroidal galaxies
to obtain an estimate of their $w(\theta)$. The SCUBA-2 Wide-area
Extragalactic survey will be much better suited for this purpose
and may indeed yield more accurate information on all three
clustering regimes than the other surveys discussed here. It will
also detect hundreds of strongly lensed proto-spheroids; these
however will be outnumbered by a factor $\simeq 100$ by unlensed
proto-spheroids, so that singling them out will require a toilsome
effort.

The experience made in different contexts (e.g. analyses of
Spitzer low-frequency data -- Frayer et al. 2006a,b -- or of WMAP
data -- Lopez-Caniego et al. 2007) has demonstrated that source
detection algorithms, when applied to low-resolution maps, easily
miss real sources or attribute to them wrong positions or
misinterpret fluctuations of Galactic foregrounds or confusion
noise in overdense regions as real sources. This implies that the
subtraction of sources above the nominal detection limit leaves
substantial residuals, that, in the case of CMB experiments, add
to the contamination of CMB maps. The situation improves
substantially if we can take advantage of the prior knowledge of
source positions. This means that the SASSy and a very large area
{\sc Herschel} survey is extremely beneficial for detection and
photometry of {\sc Planck} sources.

For the purpose of cleaning {\sc Planck} CMB maps from point
source contamination, it should be noted that, as illustrated by
the $850\,\mu$m panels of Fig.~\ref{powerspHFI}, point sources are
important on small angular scales (large $\ell$). Thus, although
the SASSy survey will be an exceedingly important resource, a very
large area {\sc Herschel}/SPIRE survey will bring in key
additional information. In fact, thanks to the steep rise of dust
emission spectra with increasing frequency, a $250\,\mu$m survey
to a flux limit of 100 mJy is deep enough to observationally
characterize both Poisson and clustering fluctuations due to all
the main populations of dusty galaxies, providing key input
information for techniques, like the widely used Maximum Entropy
Method, that need knowledge of the power spectra of the various
components to allow an accurate subtraction of their contribution
from CMB maps.

The considered surveys will also provide the first complete,
sub-mm selected samples of rare objects, such as blazars. Owing to
their much flatter counts (compared to those of dusty galaxies),
blazars are prominent in shallow surveys, while they are swamped
by dusty galaxies at faint flux levels. We thus expect that the
richest blazar sample ($\sim 1200$ objects) will come from the
SCUBA-2 SASSy survey. The {\sc Planck} $850\,\mu$m survey should
detect $\sim 550$ blazars at $|b| > 30^\circ$, while the number of
detections is expected to rapidly decrease in shorter wavelength
channels, as a consequence of the rapidly increasing detection
limit. Although {\sc Herschel}/SPIRE surveys of $\sim
400\,\hbox{deg}^2$ to a flux limit of $\sim 100\,$mJy may detect
only $\simeq 30$ blazars, the multi-wavelength information would
be of great value to investigate the currently controversial issue
of the ``blazar sequence''.

\section*{ACKNOWLEDGMENTS}
We are grateful to the referee for comments that helped
substantially improving the paper. Work supported in part by MIUR
and ASI.


\begin{thebibliography}{}

\bibitem[]{} Alexander, D.M., et al., 2003, AJ, 125, 383

\bibitem[]{} Alexander, D.M., Smail, I., Bauer, F.E., Chapman, S.C.,
Blain, A.W., Brandt, W.N., Ivison, R.J., 2005, Natur, 434, 738


\bibitem[]{} Ant{\' o}n S., Browne I.W.A., 2005, MNRAS, 356, 225

\bibitem[\protect\citeauthoryear{Aretxaga et
al.}{2007}]{2007astro.ph..2503A} Aretxaga I., et al., 2007, MNRAS,
submitted, arXiv:astro-ph/0702503

\bibitem[]{} Baugh C.M., Cole S., Frenk C.S., 1996, MNRAS, 282, 27

\bibitem[\protect\citeauthoryear{Baugh et al.}{2005}]{2005MNRAS.356.1191B}
Baugh C.M., Lacey C.G., Frenk C.S., Granato G.L., Silva L.,
Bressan A., Benson A.J., Cole S., 2005, MNRAS, 356, 1191

\bibitem[]{} Becker R.H., White R.L., Helfand D.J., 1995, ApJ, 450, 559

\bibitem[]{} Blain A.W., 1996, MNRAS, 283, 1340

\bibitem[]{} Blain A.W., Chapman S.C., Smail I., Ivison R., 2004, ApJ, 611, 725

\bibitem[\protect\citename{Blain}2005]{blain}
Blain A.W., Chapman S.C., Smail I., Ivison R., 2005, proc. ESO
workshop on "Multiwavelength Mapping of Galaxy Formation and
Evolution", p. 94

\bibitem[\protect\citeauthoryear{Blain \&
Longair}{1993}]{1993MNRAS.264..509B} Blain A.W., Longair M.S.,
1993, MNRAS, 264, 509

\bibitem[\protect\citeauthoryear{Blake, Mauch, \&
Sadler}{2004}]{2004MNRAS.347..787B} Blake C., Mauch T., Sadler
E.~M., 2004, MNRAS, 347, 787

\bibitem[\protect\citeauthoryear{Blake \& Wall}{2002}]{2002MNRAS.329L..37B}
Blake C., Wall J., 2002a, MNRAS, 329, L37

\bibitem[\protect\citeauthoryear{Blake \& Wall}{2002}]{2002MNRAS.337..993B}
Blake C., Wall J., 2002b, MNRAS, 337, 993

\bibitem[]{} Borys C., et al., 2004, MNRAS, 352, 759


\bibitem[\protect\citeauthoryear{Browne et al.}{2003}]{2003MNRAS.341...13B}
Browne I.W.A., et al., 2003, MNRAS, 341, 13


\bibitem[\protect\citename{Caputi}2006]{Caputi}
Caputi K.I., et al. 2006, ApJ, 637, 727



\bibitem[]{} Chae K., Chen G., Ratra B., Lee D., 2004, ApJ, 607, L71


\bibitem[]{} Chapman S.C., Blain A.W., Smail I., Ivison R.J., 2005, ApJ, 622, 772



\bibitem[\protect\citeauthoryear{Chary et al.}{2004}]{2004ApJS..154...80C}
Chary R., et al., 2004, ApJS, 154, 80

\bibitem[\protect\citeauthoryear{Cirasuolo et
al.}{2005}]{2005ApJ...629..816C} Cirasuolo M., Shankar F., Granato
G.L., De Zotti G., Danese L., 2005, ApJ, 629, 816

\bibitem[]{} Condon J.J., Cotton W.D., Greisen E.W., Yin Q.F.,
Perley R.A., Taylor G.B., Broderick J.J., 1998 AJ,  115, 1693

\bibitem[\protect\citeauthoryear{Cooray \&
Sheth}{2002}]{2002PhR...372....1C} Cooray A., Sheth R., 2002, PhR,
372, 1

\bibitem[\protect\citeauthoryear{Coppin et al.}{2006}]{2006MNRAS.372.1621C}
Coppin K., et al., 2006, MNRAS, 372, 1621

\bibitem[\protect\citeauthoryear{Costamante \&
Ghisellini}{2002}]{2002A&A...384...56C} Costamante L., Ghisellini
G., 2002, A\&A, 384, 56

\bibitem[\protect\citeauthoryear{Croom et al.}{2005}]{2005MNRAS.356..415C}
Croom S.M., et al., 2005, MNRAS, 356, 415

\bibitem[\protect\citeauthoryear{Daddi et al.}{2001}]{2001A&A...376..825D}
Daddi E., Broadhurst T., Zamorani G., Cimatti A., R{\"o}ttgering
H., Renzini A., 2001, A\&A, 376, 825

\bibitem[\protect\citeauthoryear{Daddi et al.}{2003}]{2003ApJ...588...50D}
Daddi E., et al., 2003, ApJ, 588, 50



\bibitem[\protect\citeauthoryear{de Zotti et
al.}{1996}]{1996ApL&C..35..289D} De Zotti G., Franceschini A.,
Toffolatti L., Mazzei P., Danese L., 1996, ApL\&C, 35, 289

\bibitem[]{} De Zotti G., Ricci R., Mesa D., Silva L., Mazzotta P.,
Toffolatti L., Gonz{\' a}lez-Nuevo J., 2005, A\&A, 431, 893

\bibitem[\protect\citeauthoryear{Dole et al.}{2004}]{2004ApJS..154...87D}
Dole H., et al., 2004, ApJS, 154, 87

\bibitem[\protect\citeauthoryear{Donato et al.}{2001}]{2001A&A...375..739D}
Donato D., Ghisellini G., Tagliaferri G., Fossati G., 2001, A\&A,
375, 739

\bibitem[\protect\citeauthoryear{Dunlop et al.}{2004}]{2004MNRAS.350..769D}
Dunlop J.~S., et al., 2004, MNRAS, 350, 769

\bibitem[]{} Dunne L., Eales S.A., 2001, MNRAS, 327, 697

\bibitem[\protect\citeauthoryear{Dunne et al.}{2000}]{2000MNRAS.315..115D}
Dunne L., Eales S., Edmunds M., Ivison R., Alexander P., Clements
D.L., 2000, MNRAS, 315, 115

\bibitem[\protect\citeauthoryear{Eisenstein \&
Hu}{1998}]{1998ApJ...496..605E} Eisenstein D.J., Hu W., 1998, ApJ,
496, 605

\bibitem[\protect\citename{fadda06}2006]{fadda06}
Fadda D. et al. 2006, astro-ph/0603488

\bibitem[\protect\citeauthoryear{Farrah et al.}{2006a}]{2006ApJ...641L..17F}
Farrah D., et al., 2006a, ApJ, 641, L17

\bibitem[\protect\citeauthoryear{Farrah et
al.}{2006b}]{2006ApJ...643L.139F} Farrah D., et al., 2006b, ApJ,
643, L139

\bibitem[\protect\citeauthoryear{Finkbeiner, Davis, \&
Schlegel}{1999}]{1999ApJ...524..867F} Finkbeiner D.P., Davis M.,
Schlegel D.J., 1999, ApJ, 524, 867

\bibitem[]{} Fossati G., Maraschi L., Celotti A., Comastri A.,
Ghisellini G., 1998, MNRAS, 299, 433

\bibitem[\protect\citeauthoryear{Frayer et al.}{2006}]{2006AJ....131..250F}
Frayer D.~T., et al., 2006a, AJ, 131, 250

\bibitem[\protect\citeauthoryear{Frayer et al.}{2006}]{2006ApJ...647L...9F}
Frayer D.~T., et al., 2006b, ApJ, 647, L9

\bibitem[\protect\citeauthoryear{Ghisellini et
al.}{1998}]{1998MNRAS.301..451G} Ghisellini G., Celotti A.,
Fossati G., Maraschi L., Comastri A., 1998, MNRAS, 301, 451


\bibitem[]{} Granato G.L., De Zotti G., Silva L., Bressan A., Danese L., 2004, ApJ, 600,
580

\bibitem[\protect\citeauthoryear{Granato et
al.}{2006}]{2006MNRAS.368L..72G} Granato G.~L., Silva L., Lapi A.,
Shankar F., De Zotti G., Danese L., 2006, MNRAS, 368, L72


\bibitem[]{} Granato G.L., Silva L., Monaco P., Panuzzo P., Salucci P., De Zotti G., Danese
L., 2001, MNRAS, 324, 757

\bibitem[]{} Gregory P.C., Scott W.K., Douglas K., Condon J.J.,
1996, ApJS,  103, 427



\bibitem[\protect\citeauthoryear{Griffith \&
Wright}{1993}]{1993AJ....105.1666G} Griffith M.R., Wright A.E.,
1993, AJ, 105, 1666

\bibitem[\protect\citeauthoryear{Gruppioni et
al.}{2002}]{2002MNRAS.335..831G} Gruppioni C., et al., 2002,
MNRAS, 335, 831

\bibitem[\protect\citeauthoryear{Harwit}{2004}]{2004AdSpR..34..568H} Harwit
M., 2004, AdSpR, 34, 568

\bibitem[\protect\citeauthoryear{Hughes \&
Gazta{\~n}aga}{2000}]{2000sfsl.conf...29H} Hughes D.H.,
Gazta{\~n}aga E., 2000, proc. ESLAB symp. Star formation from the
small to the large scale, F. Favata, A. Kaas, and A. Wilson eds.,
ESA SP 445, p. 29


\bibitem[\protect\citeauthoryear{Kaiser}{1984}]{1984ApJ...284L...9K} Kaiser
N., 1984, ApJ, 284, L9

\bibitem[\protect\citeauthoryear{Knapp \&
Patten}{1991}]{1991AJ....101.1609K} Knapp G.~R., Patten B.~M.,
1991, AJ, 101, 1609

\bibitem[]{} Kneib J., van der Werf P.P., Kraiberg Knudsen K., Smail I.,
Blain A., Frayer D., Barnard V., Ivison R., 2004, MNRAS, 349, 1211

\bibitem[\protect\citeauthoryear{Knox}{1995}]{1995PhRvD..52.4307K} Knox L.,
1995, PhRvD, 52, 4307

\bibitem[]{} Kochanek C.S., 2004, in Kochanek C.S., Schneider
P., Wambsganss J., Part 2 of Gravitational Lensing: Strong, Weak
\& Micro, Proceedings of the 33rd Saas-Fee Advanced Course, G.
Meylan, P. Jetzer \& P. North, eds. (Springer-Verlag: Berlin),
astro-ph/0407232

\bibitem[]{} Kuhlen M., Keeton C.R., Madau P., 2004, ApJ, 601, 104

\bibitem[\protect\citeauthoryear{Lagache, Dole, \&
Puget}{2003}]{2003MNRAS.338..555L} Lagache G., Dole H., Puget
J.-L., 2003, MNRAS, 338, 555

\bibitem[\protect\citeauthoryear{Lagache et
al.}{2004}]{2004ApJS..154..112L} Lagache G., et al., 2004, ApJS,
154, 112

\bibitem[\protect\citeauthoryear{Lagache, Puget, \&
Dole}{2005}]{2005ARA&A..43..727L} Lagache G., Puget J.-L., Dole
H., 2005, ARA\&A, 43, 727

\bibitem[\protect\citeauthoryear{Lamarre et
al.}{2003}]{2003NewAR..47.1017L} Lamarre J.M., et al., 2003,
NewAR, 47, 1017

\bibitem[]{} Lapi A., Shankar F., Mao J., Granato G.L., Silva L.,
De Zotti G., Danese L., 2006, ApJ, accepted, astro-ph/0603819

\bibitem[\protect\citeauthoryear{L{\'o}pez-Caniego et
al.}{2006}]{2006MNRAS.tmp..743L} L{\'o}pez-Caniego M., Herranz D.,
Gonz{\'a}lez-Nuevo J., Sanz J.L., Barreiro R.B., Vielva P.,
Arg{\"u}eso F., Toffolatti L., 2006, MNRAS, 370, 2047

\bibitem[]{} L\'opez-Caniego M., Gonz\'alez-Nuevo J., Herranz D., Massardi
M., Sanz J.L., De Zotti G., Toffolatti L., Arg\"ueso F., 2007,
ApJS, accepted

\bibitem[\protect\citeauthoryear{Madgwick et
al.}{2003}]{2003MNRAS.344..847M} Madgwick D.S., et al., 2003,
MNRAS, 344, 847


\bibitem[\protect\citeauthoryear{Magliocchetti et
al.}{1999}]{1999MNRAS.306..943M} Magliocchetti M., Maddox S.J.,
Lahav O., Wall J.V., 1999, MNRAS, 306, 943

\bibitem[]{} Magliocchetti M., Silva L., Lapi A., De
Zotti G., Granato G.L., Fadda D., Danese L., 2006, MNRAS,
accepted, astro-ph/0611409

\bibitem[\protect\citeauthoryear{Marinoni \&
Hudson}{2002}]{2002ApJ...569..101M} Marinoni C., Hudson M.J.,
2002, ApJ, 569, 101

\bibitem[\protect\citeauthoryear{Matteucci}{2006}]{2006AIPC..847...79M}
Matteucci F., 2006, AIPC, 847, 79

\bibitem[\protect\citeauthoryear{Mauch et al.}{2003}]{2003MNRAS.342.1117M}
Mauch T., Murphy T., Buttery H.J., Curran J., Hunstead R.W.,
Piestrzynski B., Robertson J.G., Sadler E.~M., 2003, MNRAS, 342,
1117

\bibitem[\protect\citeauthoryear{Mazzei, Xu, \& de
Zotti}{1992}]{1992A&A...256...45M} Mazzei P., Xu C., de Zotti G.,
1992, A\&A, 256, 45


\bibitem[\protect\citeauthoryear{McKay et al.}{2002}]{2002ApJ...571L..85M}
McKay T.A., et al., 2002, ApJ, 571, L85


\bibitem[]{} Mitchell J.L., Keeton C.R., Frieman J.A., Sheth R.K., 2005, ApJ, 622,
81

\bibitem[\protect\citeauthoryear{Mo \& White}{1996}]{1996MNRAS.282..347M}
Mo H.J., White S.D.M., 1996, MNRAS, 282, 347

\bibitem[\protect\citeauthoryear{Mortier et al.}{2005}]{2005MNRAS.363..563M}
Mortier A.M.J., et al., 2005, MNRAS, 363, 563


\bibitem[\protect\citeauthoryear{Moustakas \&
Somerville}{2002}]{2002ApJ...577....1M} Moustakas L.A., Somerville
R.S., 2002, ApJ, 577, 1

\bibitem[]{} Myers S.T., et al., 2003, MNRAS, 341, 1

\bibitem[]{} Negrello M., Magliocchetti M., Moscardini L., De Zotti G.,
Granato G.L., Silva L., 2004, MNRAS, 352, 493

\bibitem[\protect\citeauthoryear{Negrello, Magliocchetti, \& De
Zotti}{2006}]{2006MNRAS.368..935N} Negrello M., Magliocchetti M.,
De Zotti G., 2006, MNRAS, 368, 935

\bibitem[\protect\citeauthoryear{Nieppola, Tornikoski, \&
Valtaoja}{2006}]{2006A&A...445..441N} Nieppola E., Tornikoski M.,
Valtaoja E., 2006, A\&A, 445, 441


\bibitem[\protect\citeauthoryear{Overzier et
al.}{2003}]{2003A&A...405...53O} Overzier R.A., R{\"o}ttgering
H.J.A., Rengelink R.B., Wilman R.J., 2003, A\&A, 405, 53

\bibitem[\protect\citeauthoryear{Papovich et
al.}{2004}]{2004ApJS..154...70P} Papovich C., et al., 2004, ApJS,
154, 70


\bibitem[\protect\citeauthoryear{Pearson et
al.}{2004}]{2004MNRAS.347.1113P} Pearson C.P., et al., 2004,
MNRAS, 347, 1113

\bibitem[\protect\citeauthoryear{Peebles}{1980}]{1980lssu.book.....P}
Peebles P.J.E., 1980, The large-scale structure of the universe,
Princeton University Press

\bibitem[\protect\citeauthoryear{Percival et
al.}{2003}]{2003MNRAS.338L..31P} Percival W.J., Scott D., Peacock
J.A., Dunlop J.S., 2003, MNRAS, 338, L31

\bibitem[\protect\citeauthoryear{P{\'e}rez-Gonz{\'a}lez et
al.}{2005}]{2005ApJ...630...82P} P{\'e}rez-Gonz{\'a}lez P.G., et
al., 2005, ApJ, 630, 82

\bibitem[]{} Perrotta F., Baccigalupi C., Bartelmann M., De Zotti G., Granato
G.~L., 2002, MNRAS, 329, 445

\bibitem[]{} Perrotta F., Magliocchetti M., Baccigalupi C., Bartelmann M., De Zotti G.,
Granato G.~L., Silva L., Danese L., 2003, MNRAS, 338, 623

\bibitem[\protect\citeauthoryear{Porciani, Magliocchetti, \&
Norberg}{2004}]{2004MNRAS.355.1010P} Porciani C., Magliocchetti
M., Norberg P., 2004, MNRAS, 355, 1010



\bibitem[\protect\citeauthoryear{Ricci et al.}{2004}]{ricci04}
  Ricci R., et al., 2004, MNRAS, 354, 305

\bibitem[\protect\citeauthoryear{Ricci et al.}{2006}]{2006A&A...445..465R}
Ricci R., Prandoni I., Gruppioni C., Sault R.~J., de Zotti G.,
2006, A\&A, 445, 465


\bibitem[\protect\citeauthoryear{Saunders et
al.}{2000}]{2000MNRAS.317...55S} Saunders W., et al., 2000, MNRAS,
317, 55

\bibitem[\protect\citeauthoryear{Saunders et
al.}{1990}]{1990MNRAS.242..318S} Saunders W., Rowan-Robinson M.,
Lawrence A., Efstathiou G., Kaiser N., Ellis R.S., Frenk C.S.,
1990, MNRAS, 242, 318

\bibitem[\protect\citeauthoryear{Scott, Dunlop, \&
Serjeant}{2006}]{2006MNRAS.370.1057S} Scott S.E., Dunlop J.S.,
Serjeant S., 2006, MNRAS, 370, 1057

\bibitem[]{} Scott S.E., et al., 2002, MNRAS, 331, 817

\bibitem[]{} Serjeant S., Harrison D., 2005, MNRAS, 356, 192

\bibitem[\protect\citeauthoryear{Shankar et
al.}{2006}]{2006ApJ...643...14S} Shankar F., Lapi A., Salucci P.,
De Zotti G., Danese L., 2006, ApJ, 643, 14

\bibitem[\protect\citeauthoryear{Sheth \&
Tormen}{1999}]{1999MNRAS.308..119S} Sheth R.K., Tormen G., 1999,
MNRAS, 308, 119

\bibitem[]{} Silva L., De Zotti G., Granato G.L., Maiolino R., Danese L., 2004, astro-ph/0403166

\bibitem[\protect\citeauthoryear{Silva et al.}{2005}]{2005MNRAS.357.1295S}
Silva L., De Zotti G., Granato G.~L., Maiolino R., Danese L.,
2005, MNRAS, 357, 1295

\bibitem[\protect\citeauthoryear{Smail et al.}{2003}]{2003ApJ...599...86S}
Smail I., Scharf C.A., Ivison R.J., Stevens J.A., Bower R.G.,
Dunlop J.S., 2003, ApJ, 599, 86



\bibitem[\protect\citeauthoryear{van Kampen et
al.}{2005}]{2005MNRAS.359..469V} van Kampen E., et al., 2005,
MNRAS, 359, 469

\bibitem[]{} Vielva P., Mart{\'{\i}}nez-Gonz{\' a}lez E., Gallegos J.E., Toffolatti
L., Sanz J.L., 2003, MNRAS, 344, 89

\bibitem[Waldram et al.(2003)]{waldram03} Waldram E.M., Pooley G.G.,
  Grainge K., Jones M.E., Saunders R.D.E., Scott P.F. \& Taylor A.C.,
  2003, MNRAS, 342, 915

\bibitem[\protect\citeauthoryear{Webb et al.}{2003}]{2003ApJ...587...41W}
Webb T.M., et al., 2003, ApJ, 587, 41

\bibitem[]{} White R.L., Becker R.H., Helfand D.J., Gregg M.D., 1997, ApJ, 475, 479


\end{thebibliography}
\end{document}